\documentclass[letterpaper,times]{IONconf}

\usepackage{amsmath}
\usepackage{amssymb}

\usepackage{graphicx}
\usepackage[justification=centering]{caption}

\usepackage{tikz}
\usetikzlibrary{calc,arrows.meta}

\usepackage{url}

\usepackage{natbib}

\usepackage[hidelinks]{hyperref}
\usepackage{gensymb}
\usepackage{booktabs}

\usepackage[skip=2pt]{caption}  
\usepackage{float}               

\title{Space-Based GNSS Radio Frequency Interference Detection Evaluation Through Multi-Satellite Data Integration}

\author{
    Anouar~Boumeftah, \textit{Polytechnique~Montréal}%
    \vspace{1mm} \\%
    Peter Klimas, \textit{Northstar Earth \& Space}%
    \vspace{1mm} \\%
    Gunes~Karabulut~Kurt, \textit{Polytechnique~Montréal}%
}

\begin{document}

\maketitle

\section*{Biography}
\biography{Anouar Boumeftah}{is a graduate student in aerospace engineering at Polytechnique Montréal, where his research focuses on satellite security and resilience. He is also an Advanced Member of Technical Staff at NorthStar Earth \& Space, contributing to operational Space Situational Awareness systems.}

\biography{Peter Klimas}{is the Executive Director of Engineering at NorthStar Earth \& Space, leading engineering strategy and development of space-based Space Situational Awareness systems. He received undergraduate and graduate degrees in Mechanical and Aerospace Engineering from Carleton University.}

\biography{Gunes Karabulut Kurt}{is a Canada Research Chair (Tier 1) in New Frontiers in Space Communications and Full Professor at Polytechnique Montréal. She is also an Adjunct Research Professor at Carleton University, with research interests in space security, multi-functional space networks, and wireless communications.}


\section*{Abstract}
Space-based GNSS reflectometry (GNSS-R) can detect terrestrial radio frequency interference (RFI) through elevated noise power in delay-Doppler map forbidden zones. This study evaluates how constellation size affects detection performance using Level 1 delay-Doppler observations from seven CYGNSS spacecraft collected over three months from the NASA PO.DAAC archive. Four metrics are analysed: detection latency, spatial coverage, spatial coherence, and persistence monitoring reliability. Results show that the full seven-satellite constellation reduces median detection latency by a factor of 4.7 compared with a single satellite and increases interception probability for a 5-minute emission from 2\% to 11.5\%. Median footprint revisit time improves from 5.8 hours to under 2.0 hours. Spatial coherence analysis indicates that a single satellite leaves up to 72\% of source structure unresolved. Persistence monitoring confirms interference onset 39 days earlier than single-satellite deployment. The largest gains occur between one and three satellites, establishing three satellites as the minimum effective constellation size.

\section{Introduction}

Global navigation satellite systems (GNSS) have become essential infrastructure for positioning, navigation, and timing (PNT) services across civilian, scientific, and military domains. Because GNSS signals are received at very low power levels, they remain inherently vulnerable to radio frequency interference (RFI), including both intentional and unintentional sources (\cite{dovis2015}). This vulnerability creates growing risks for safety-critical and infrastructure-dependent applications such as aviation, telecommunications, and remote sensing. Space-based monitoring has emerged as a promising complement to ground-based GNSS RFI surveillance. In particular, Low Earth Orbit (LEO) platforms carrying GNSS receivers and GNSS-reflectometry (GNSS-R) payloads can provide broad-area coverage and detect terrestrial interference through observables such as elevated delay-Doppler map (DDM) noise floors (\cite{roberts2022detection, chew2023rfi}). These capabilities make multi-satellite space-based architectures attractive for large-scale and persistent RFI monitoring.

However, while prior work has demonstrated that space-based GNSS instruments can detect terrestrial interference, the specific performance gains obtained by integrating observations from multiple satellites have not been systematically quantified. For operational monitoring architectures, this distinction is critical: the value of a constellation lies not only in increased data availability, but also in improvements in detection latency, spatial-temporal coverage, revisit behavior, and persistent monitoring capability compared to single-satellite systems.

This paper addresses that gap by evaluating GNSS RFI detection performance through multi-satellite data integration using real observations. The main contributions of this work are as follows:

\begin{enumerate}
    \item We develop a quantitative evaluation framework to compare single-satellite and multi-satellite GNSS-R RFI detection performance using real CYGNSS observations.
    
    \item We characterize constellation-level performance improvements across key operational dimensions, including detection latency, probability of intercept for transient interference events, spatial coverage, revisit behavior, and persistence monitoring capability.

    \item We design and implement an automated data ingestion and processing pipeline that (1) retrieves CYGNSS Level~1 data directly from the NASA PO.DAAC OPeNDAP interface (\cite{cygnss_l1}), (2) performs data quality filtering and geometric corrections, and (3) generates analysis-ready datasets for large-scale RFI monitoring studies.

    \item We develop an interactive geospatial visualization and analysis environment that enables exploration of detected RFI activity, constellation coverage patterns, and temporal interference behavior derived from the processing framework.
\end{enumerate}

The remainder of this paper is organized as follows: Section~\ref{sec:litreview} reviews relevant literature on space-based GNSS RFI detection and multi-satellite remote sensing. Section~\ref{sec:cygnss} describes the CYGNSS mission and data characteristics. Section~\ref{sec:methodology} presents the detection methodology and comparison framework. Sections~\ref{sec:results} and~\ref{sec:discussion} present the results and discussion, and Section~\ref{sec:conclusion} concludes the paper.


\section{Literature Review}
\label{sec:litreview}

GNSS signals operate in the L-band with received power levels at Earth's surface typically around $-130$ dBm for GPS L1 C/A signals, making them inherently vulnerable to RFI (\cite{dovis2015}). RFI sources include both unintentional interference from legitimate services and intentional attacks such as jamming and spoofing. Recent comprehensive reviews document the proliferation of detection methods, with increasing emphasis on machine learning approaches for automated classification (\cite{brkic2024recent, wu2020spoofing}). The availability of low-cost software-defined radios has increased access to interference generation capabilities, raising concerns about critical infrastructure protection (\cite{humphreys2017}).

Ground-based RFI monitoring using networks of GNSS receivers has demonstrated effectiveness for regional coverage. \cite{contreras2023lowcost} showed that dense receiver networks can significantly improve probability of intercept for interference signals, particularly for transient or low-duty-cycle emitters. However, ground networks face inherent limitations in geographic coverage, deployment costs in remote regions, and political access constraints in conflict zones where RFI activity is often concentrated.

Space-based platforms offer complementary capabilities addressing ground-based limitations. Constellation of LEO satellites equipped with GNSS receivers provide global or near-global coverage with unique geometric perspectives unavailable to terrestrial systems. \cite{bonnedal2010radio} pioneered space-based RFI detection using the GRAS instrument aboard MetOp satellites. By monitoring automatic gain control and tracking high-gain radio occultation antennas during limb-viewing geometry, they can detect elevated noise power indicative of surface-based jammers.

\cite{roberts2022detection} developed highly sensitive detection algorithms using nearly 20 years of data from NASA's Blackjack/TriG GNSS receivers aboard COSMIC and GRACE satellites. Their method identified RFI events by detecting correlated signal-to-noise ratio variations across multiple simultaneously tracked satellites, a signature characteristic of terrestrial interference. They created global annual maps revealing RFI hotspots with strong correlation to regional conflicts and geopolitical activity. \cite{murrian2021detection} achieved precise geolocation of interference sources using a GNSS receiver aboard the ISS, demonstrating continuous operation of certain jammers since 2017. \cite{wu2023gnss} analyzed GNSS radio occultation data from multiple LEO missions including MetOp and COSMIC-2, developing algorithms to detect and monitor jamming signals. Their work revealed temporal evolution of RFI hotspots and demonstrated that deeply-occulted observations provide particularly sensitive indicators of ground-based interference.

Understanding GNSS signal reflections from the Earth’s surface is critical for interpreting variations in received signals, which can affect the detection and characterization of terrestrial interference. GNSS-R uses downward-looking high-gain antennas to receive surface-reflected GNSS signals for Earth remote sensing (\cite{zavorotny2014tutorial}). The fundamental GNSS-R observable is the delay-Doppler map (DDM), generated by cross-correlating received signals with replicas over a range of delays and Doppler frequencies (\cite{zavorotny2000scattering}). The Zavorotny-Voronovich model provides the theoretical foundation for DDM physics, describing scattered signal power as a function of geometric and environmental parameters.

The downward-looking antenna configuration employed by GNSS-R missions makes these systems exceptionally sensitive to ground-based RFI. \cite{chew2023rfi} demonstrated that RFI manifests as elevated noise power levels in the forbidden zone of DDMs, the early delay bins where no legitimate reflected signal should appear. Analyzing CYGNSS Level 1 data from the eight-satellite constellation from 2017 to 2022, they identified numerous RFI hotspots and quantified transmission durations, confirming strong correlations between RFI appearance and geopolitical conflicts. They also addressed the challenge that onboard kurtosis-based RFI flags exhibit high false alarm rates over land due to terrain and inland water effects. They proposed using mean noise power in DDM forbidden zones as a more robust metric, providing the methodological foundation adopted in subsequent studies (\cite{wu2022detection}).

The remote sensing literature extensively documents benefits of multi-platform approaches, though largely focused on combining different sensor types rather than identical sensors on multiple platforms. A single LEO satellite in a 500-km orbit typically revisits a given location every 12 to 24 hours depending on inclination and latitude. Multi-satellite constellations can dramatically reduce revisit times through coordinated coverage (\cite{ruf2016new}). \cite{alkhaldi2025} exploited both Spire and CYGNSS products to show clear evidence of RFI contributions. The FSSCat CubeSat mission (\cite{camps2020fsscat}) demonstrated dual-purpose payloads for identifying RFI hotspots in Arctic regions, providing spatial comparison to CYGNSS's tropical and mid-latitude focus. These multi-mission comparisons established that space-based RFI detection represents a maturing operational capability across diverse orbital configurations.

Despite extensive literature on space-based RFI detection and multi-satellite remote sensing, a critical gap remains. No systematic quantitative comparison exists between single-satellite and multi-satellite constellation performance for GNSS RFI detection. Previous CYGNSS RFI studies (\cite{chew2023rfi, wu2022detection}) successfully demonstrated detection capabilities but did not isolate and quantify the specific performance gains attributable to having eight coordinated satellites versus one. This gap has practical implications for future mission design. GNSS-R missions for RFI monitoring must justify constellation size and architecture based on performance requirements. Operational systems need quantitative metrics to assess detection capabilities, latency, coverage, and reliability. ICAO and other aviation authorities require confidence bounds on probability of detection for operational decision-making (\cite{icao2025gnss}).

Our research addresses this gap through systematic performance comparison across key dimensions: detection latency quantification for new interference sources, spatial coverage and probability of intercept analysis, temporal characterization enabling automated change detection, and sensitivity enhancement through multi-platform integration. By developing comparison methodologies and applying them to real CYGNSS data covering known RFI events, we provide empirical evidence quantifying constellation benefits and informing future system design.


\section{CYGNSS Mission and Data Description}
\label{sec:cygnss}

The CYGNSS mission uses a constellation of microsatellites to provide frequent, high-resolution observations of GPS signals reflected from Earth's surface. This section summarizes the constellation design, orbital configuration, and instrument characteristics, along with key aspects of the resulting data, including spatial coverage, temporal sampling, and the effects of interference and measurement footprint.

\subsection{Constellation Architecture and Characteristics}

\begin{figure}[h]
    \centering
    \includegraphics[width=0.45\textwidth]{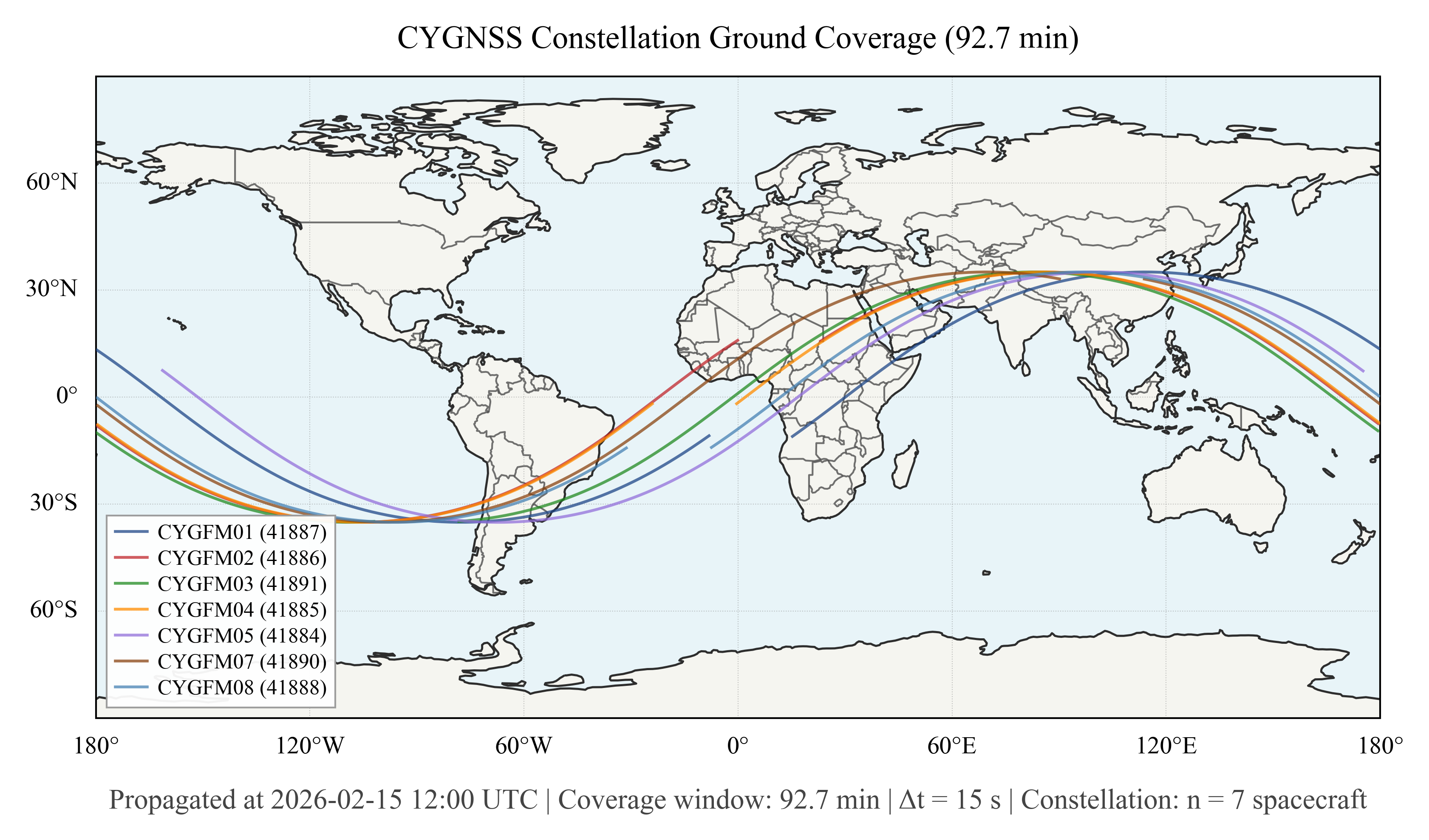}
    \includegraphics[width=0.45\textwidth]{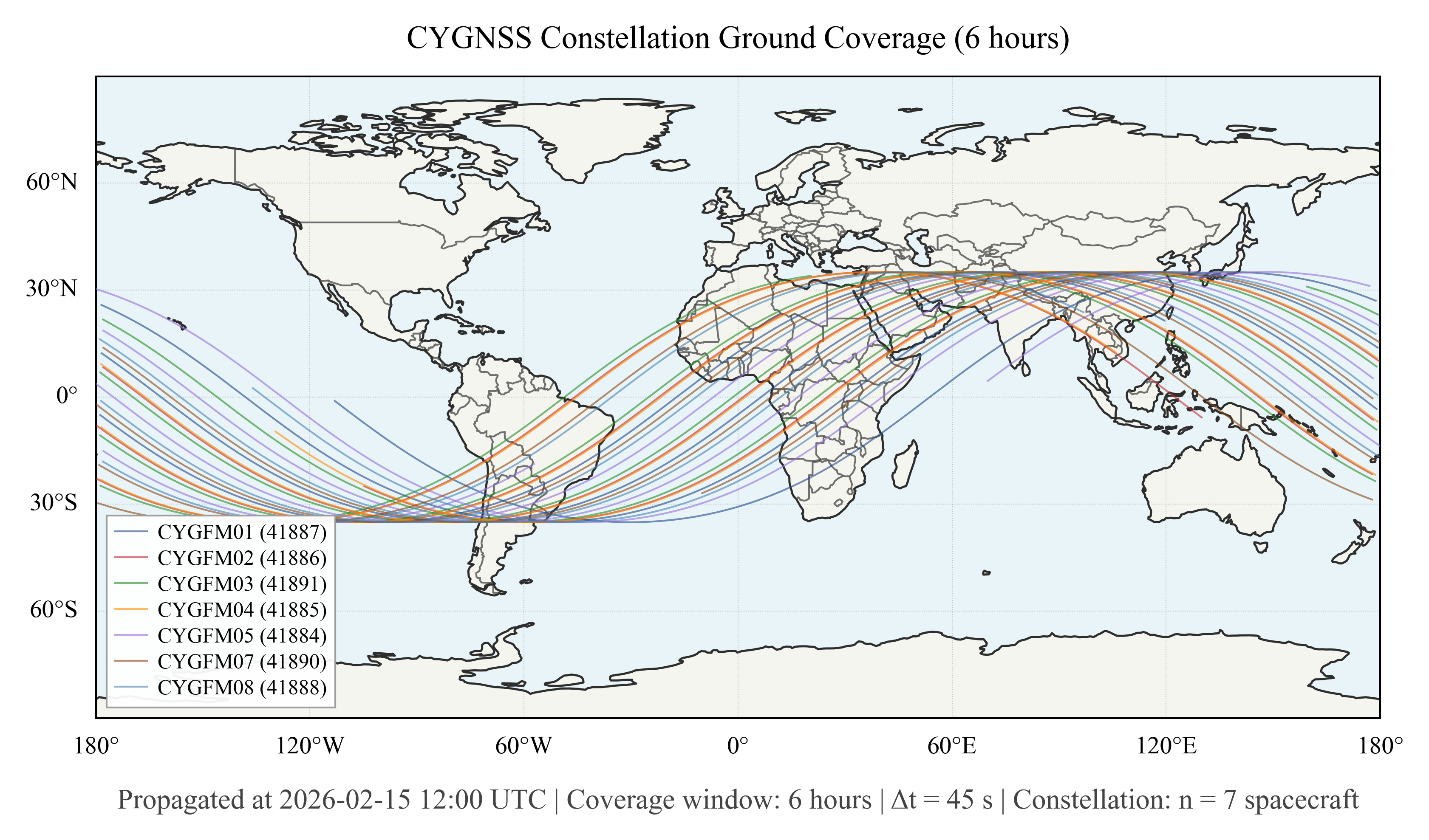}

    \vspace{0.2cm}

    \includegraphics[width=0.45\textwidth]{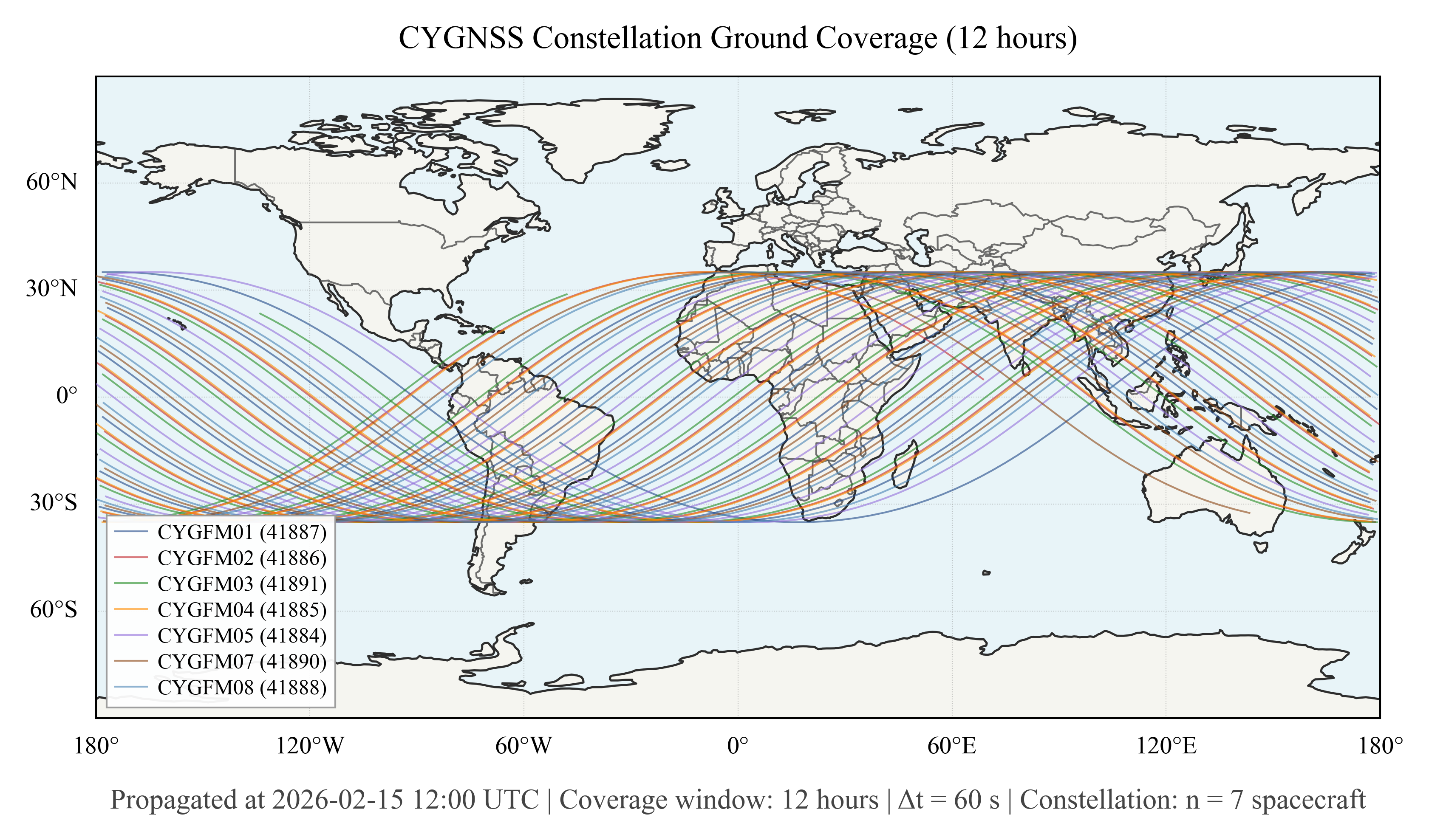}
    \includegraphics[width=0.45\textwidth]{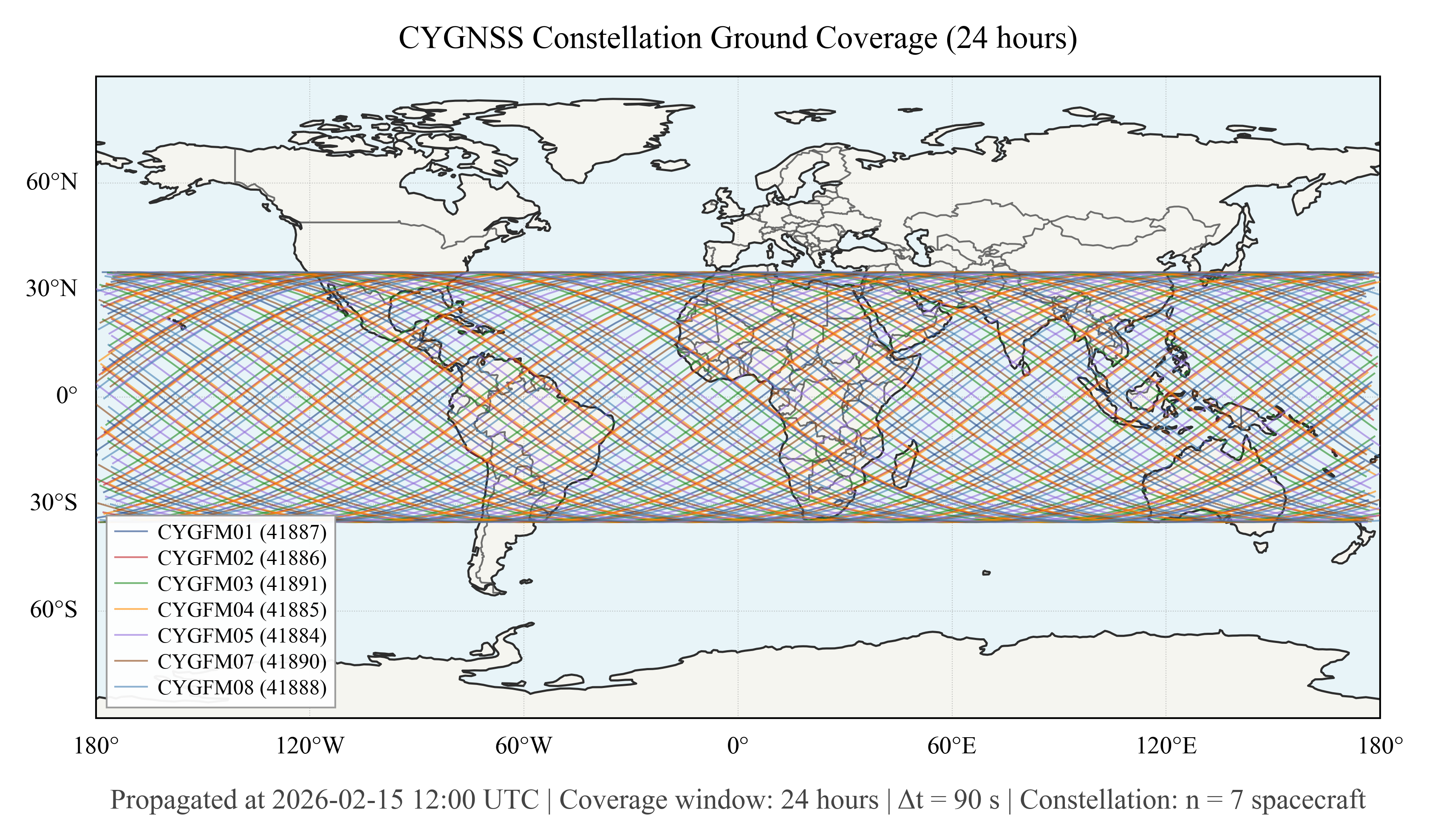}

    \caption{CYGNSS constellation ground-track evolution over increasing accumulation intervals: (a) one orbital period ($\sim$93 min), (b) 6 h, (c) 12 h, and (d) 24 h. The seven operational spacecraft progressively densify coverage within the $\pm35^\circ$ latitude band.}
    \label{fig:groundtracks}
\end{figure}

The CYGNSS constellation consists of eight microsatellites deployed into a common orbital plane at approximately 510 km altitude with $35^{\circ}$ inclination (\cite{ruf2016new, ruf2019new}). This configuration provides coverage between $\pm40^{\circ}$ latitude, encompassing tropical and mid-latitude regions. The orbital parameters yield a mean orbital period of approximately 93 minutes, with the eight satellites distributed to provide enhanced temporal sampling compared to traditional single-satellite missions. Each satellite carries two downward-looking GPS L1 antennas oriented $28^{\circ}$ off-nadir to receive surface-reflected signals for ocean wind retrieval.

Figure~\ref{fig:groundtracks} illustrates the constellation's ground coverage patterns over three temporal scales. The single-orbit view (Fig.~\ref{fig:groundtracks}a) depicts the instantaneous constellation geometry, showing the spatial distribution of the seven operational spacecraft (CYGFM01, 02, 03, 04, 05, 07, 08) during one 93-minute orbital period. The 12-hour coverage (Fig.~\ref{fig:groundtracks}b) demonstrates the revisit characteristics, with each satellite completing approximately 7--8 orbits and creating dense ground track networks across the observable latitude band. The 24-hour view (Fig.~\ref{fig:groundtracks}c) shows the complete daily coverage pattern, with 15--16 passes per satellite providing near-continuous monitoring capability within the $\pm40^{\circ}$ latitude bounds.

\begin{figure}[htbp]
\centering
\includegraphics[width=0.70\textwidth]{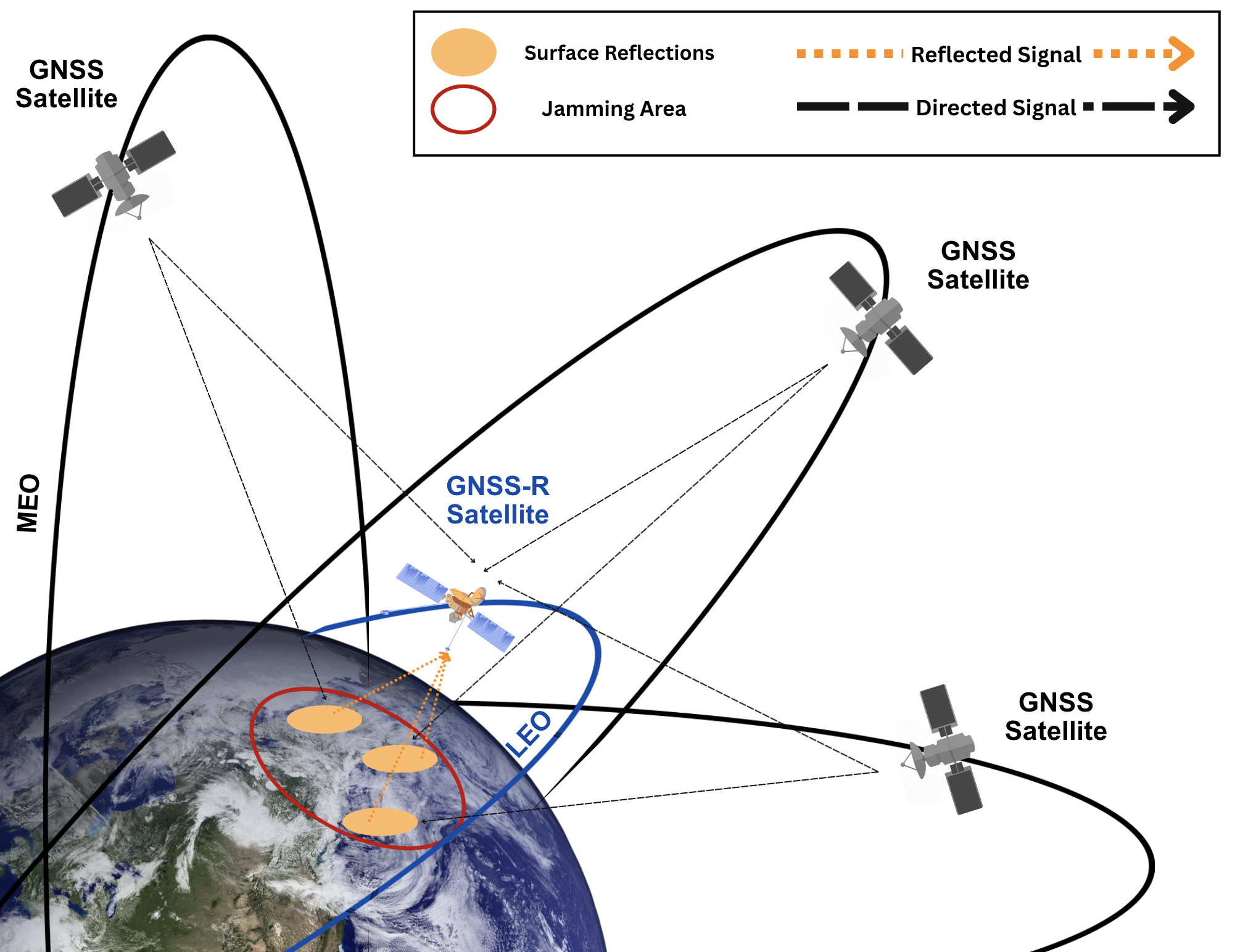}
\caption{Conceptual Diagram of GNSS Signal Reflection and Jamming Detection by a LEO GNSS-R Satellite}
\label{fig:system_model}
\end{figure}

CYGNSS exploits a bistatic radar technique that uses GPS L1 C/A signals (1575.42 MHz) transmitted from medium Earth orbit (MEO) satellites and reflected from Earth's surface (\cite{ruf2019new}). The measurement geometry is illustrated in Figure~\ref{fig:system_model}. Each CYGNSS satellite processes received signals through delay-Doppler mapping (DDM), a two-dimensional cross-correlation between the received signal and a locally generated GPS code replica across multiple time delays and Doppler frequency shifts (\cite{chew2023rfi}).

The DDM structure consists of delay bins (corresponding to range) and Doppler bins (corresponding to relative velocity between transmitter, surface, and receiver). Under nominal ocean surface conditions, the peak DDM power occurs at the specular reflection point, with additional power distributed in a characteristic horseshoe pattern due to surface roughness. The first 45 delay bins of the DDM, termed the termed the ``forbidden zone'', should theoretically contain only thermal noise as they correspond to ranges shorter than the direct path from GPS satellite to CYGNSS receiver (\cite{chew2023rfi}). This region provides the baseline for RFI detection.

The region on the Earth's surface where specular reflection points can be observed is determined by the CYGNSS antenna field of view and the geometry of GPS signal reflection. For each satellite, the two downward-looking antennas are oriented approximately $28^{\circ}$ off-nadir, with usable gain out to roughly $50^{\circ}$ incidence angle. This defines an annular region extending hundreds of kilometers from the sub-satellite track within which reflections may occur. The effective spatial resolution of individual GNSS-R measurements, however, is set by the first Fresnel zone around each specular point, which is significantly smaller at approximately $25$ km (\cite{cygnss_l1}). 
As demonstrated by \cite{chew2023rfi}, RFI shows as vertical stripes across all or most delay bins in the DDM, with power concentrated in specific Doppler bins corresponding to the relative motion between the satellite and the terrestrial emitter. The antenna inclination creates directional sensitivity: terrestrial RFI sources located north of the satellite ground track will be detected primarily by the port (north-facing) antenna, while southern sources are detected by the starboard antenna.

\section{Methodology}
\label{sec:methodology}

This section describes the methodology used in this study to process the observations and evaluate the system's capability for monitoring RFI. We first present the data processing procedures and the metrics used to identify RFI in CYGNSS measurements. We then detail our methodological framework used to assess constellation performance, including analyses of temporal detection capability, spatial coverage and footprint revisit characteristics, geometric coverage considerations, and detection and persistence monitoring of RFI activity patterns.

\subsection{Data Processing and Analysis Methods}

This study uses CYGNSS Level 1 science data (version 3.1) from calendar year 2025, accessed via NASA's Physical Oceanography Distributed Active Archive Center (PO.DAAC) (\cite{cygnss_l1}). Following \cite{chew2023rfi}, we use the mean noise power in the DDM forbidden zone (first 45 delay bins) as the primary RFI indicator, which proves more robust over land than the standard kurtosis-based flag that suffers false triggers from high-elevation terrain and coherent reflections from inland water.

We apply geometric correction for the $28^{\circ}$ antenna off-nadir pointing by computing the antenna boresight ground location rather than spacecraft nadir position, ensuring detected RFI sources align with actual terrestrial locations. All observations are projected onto a ``Plate Carrée'' projection using geographic latitude–longitude coordinates. Depending on the section of the dataset, the grid resolution varies between 1°, 0.5° and 0.1° in both latitude and longitude, with temporal binning into configurable windows depending on analysis objectives.

The methodology evaluates four dimensions of constellation performance: temporal detection, spatial coverage and revisit, geometric coherence, and persistence monitoring.

\subsubsection{Temporal Coverage: Detection Latency Analysis}
\label{sec:temporal}

This analysis evaluates constellation temporal performance across three
metrics: detection latency for persistent RFI sources, transient emission
intercept probability, and monitoring cadence through revisit intervals.
We use 48~hours of CYGNSS Level~1 DDM data (December~17--18, 2025) from
the operational 7-satellite constellation (cyg01--cyg05, cyg07, cyg08;
FM06 excluded due to an antenna anomaly). High noise floor observations
are identified using a simple threshold detector (noise floor exceeding a fixed threshold), more advanced RFI detection
methods are introduced later in Section~\ref{sec:persistence}. Spatial
binning at $1^\circ$ resolution yields 6,505 candidate hotspot cells
globally, from which cells with $\geq 10$ RFI observations are retained.
A random sample of 400 hotspots is used for detection latency and
transient analyses. 

Constellation configurations include $N=1$, $N=3$,
$N=5$, and $N=7$ satellites. To avoid orbital position bias, $N=1$
results represent the median performance across all seven satellites
tested independently, while $N=3$ and $N=5$ results report the median
across representative sub-constellation combinations.The detection criterion is based on the CYGNSS specular point footprint
radius of 527~km at the nominal 400~km
orbital altitude; an RFI source is considered observed when a satellite
specular point falls within this distance (haversine $\leq 527$~km).
Detection latency measures the elapsed time from the start of the
observation window to the first intersecting satellite pass over each
hotspot. For each configuration, intersecting observations are identified
and the earliest timestamp recorded (hours from dataset start). The
resulting cumulative distribution function (CDF) represents the fraction
of RFI sources detected within a given time threshold.

Transient RFI detection capability is evaluated via Monte Carlo simulation
with 1,000 scenarios per emission duration. Each scenario generates a
random emission with latitude sampled uniformly from $[-35^\circ,
35^\circ]$ (CYGNSS coverage band), longitude from $[-180^\circ,
180^\circ]$, start time within the 48-hour dataset range, and duration
from $\{5, 15, 30, 60, 120, 240, 480, 720, 1440\}$~minutes. Detection
occurs if any satellite observation intersects the emission's
spatiotemporal window at haversine distance $\leq 527$~km. Probability of
intercept (POI) is calculated as the fraction of detections, with 95\%
confidence intervals derived from the binomial standard error.

The revisit time analysis measures intervals between consecutive satellite
passes over the 50 highest-intensity RFI hotspots, using an extended
20-day observation window to ensure sufficient pass statistics per
location. A small search radius is applied around each hotspot,
corresponding to the DDM specular point footprint rather than the wide
swath used for detection latency. Consecutive observations separated by
less than 30~minutes are grouped into a single pass event; the revisit
gap is the elapsed time between the first observations of successive
pass events. Gaps shorter than 15~minutes are discarded as simultaneous
multi-satellite overpasses rather than true revisits. For $N=1$, only
satellites recording at least two passes over a given hotspot contribute
gaps, avoiding censoring bias from satellites whose ground tracks did not
intersect the hotspot within the window. For $N > 1$, pass timestamps
from all satellites in the sub-constellation are pooled and sorted
chronologically before differencing, so the revisit gap reflects the
waiting time until any constellation member next overflies the location.

\subsubsection{Spatial Coverage: GNSS-R Footprint Revisit Time}
\label{sec:spatial}

This analysis focuses on one particular spatial metric: GNSS-R footprint revisit time, how frequently a given location on Earth falls within the observable region where at least one satellite's downward-looking antennas can detect reflected terrestrial RFI.
Each CYGNSS satellite observes an annular footprint centered on its nadir track. With antenna boresight at $28^\circ$ off-nadir and effective gain pattern extending to incidence angles of $50^\circ$, the instantaneous observable region spans from 225~km (inner edge) to 527~km (outer edge) in the cross-track direction on each side of the ground track.

We reconstruct the full annular footprint from satellite position data at each time step rather than relying solely on specular point coordinates, which represent only individual reflection subpoints. For each satellite position, we compute the great-circle distance from the satellite's nadir to every cell in a $0.5^\circ \times 0.5^\circ$ geographic grid covering $\pm 40^\circ$ latitude.

For each constellation configuration $(N=1, N=3, N=5, N=7)$, we track cumulative coverage where a grid cell is considered covered once it falls inside at least one satellite's footprint at any time during the 48-hour analysis window. We quantify coverage fraction over time (the percentage of the $\pm 40^\circ$ latitude zone observed at least once, evaluated at 6h, 12h, 24h, and 48h milestones), revisit time distribution (for each grid cell, we record all times when it was inside a satellite footprint, then compute the mean and maximum gaps between consecutive passes), and latitude-dependent coverage statistics to characterize spatial non-uniformity arising from CYGNSS's $35^\circ$ orbital inclination.

\subsubsection{Geometric Coverage and Spatial Coherence}
\label{sec:geometric}

The preceding analyses quantify how constellation size affects detection
latency and spatial revisit. This section addresses a complementary
question: whether the noise floor elevation at a candidate RFI location
possesses the spatial structure expected of a genuine point source.
A detection is only credible if the elevated signal conforms to a
physically consistent footprint; a spatially incoherent pattern instead
suggests receiver artefacts, transient interference, or residual sampling
bias. The normalised residual variance (NRV) is used as a continuous
goodness-of-fit statistic to quantify this coherence, and its dependence
on constellation size $N$ constitutes the primary result of this section.

Before source characterisation, cells exhibiting implausible temporal
stability are masked. For each $0.1^\circ \times 0.1^\circ$ grid cell,
the inter-day standard deviation $\sigma_\mathrm{d}$ of the cell-mean
noise floor is computed; cells where $\sigma_\mathrm{d}$ falls below
$\alpha_\mathrm{hp} = 0.05$ of the regional noise floor standard deviation
are flagged as hot pixels and excluded, as their stability is physically
inconsistent with the evolving CYGNSS viewing geometry.

The spatial peak is then located on the $N=7$ accumulation map, with the
signal defined as $s_{ij} = \bar{\eta}_{ij} - \eta_\mathrm{bg}$ relative
to the mean noise floor of three quiet Southern Ocean reference regions.
An $11 \times 11$ cell smoothing filter suppresses single-cell artefacts
prior to peak detection, and only cells with at least 20 accumulated
observations are eligible as candidates. A constellation configuration
proceeds to fitting only if the mean signal elevation within the bounding
box exceeds $\Delta\eta_\mathrm{min} = 0.30$~dBW, ensuring NRV differences
across constellation sizes reflect spatial coherence rather than marginal
detectability.

For $N=7$, a full two-dimensional Gaussian is fitted to
$s_{ij}$ by nonlinear least squares with SNR-weighted residuals:
\begin{equation}
    \hat{s}(x, y) = \delta + A \exp\!\left[
        -\frac{1}{2}\!\left(
            \frac{x_r^2}{\sigma_x^2} + \frac{y_r^2}{\sigma_y^2}
        \right)
    \right],
    \label{eq:gaussian2d}
\end{equation}
where $(x_r, y_r)$ are coordinates rotated to the source major axis at
angle $\theta$, $A$ is amplitude, $\delta$ a spatially uniform offset,
and $(\sigma_x, \sigma_y)$ the semi-axis widths. Cell weights
$w_{ij} = \Delta\mathrm{SNR}_{ij} \cdot \max(s_{ij}, 0)$ prioritise
high-elevation, high-confidence observations; individual observations are
referred to by a flat index $k$ in subsequent expressions. For $N \in \{1, 3, 5\}$,
shape parameters are locked to the $N=7$ template and only amplitude and
offset are re-estimated by weighted linear least squares, so that NRV
measures strictly how well the sparser constellation's data conforms to
the spatial structure established by the full fleet.

NRV is defined as the weighted ratio of residual to signal variance:
\begin{equation}
    \mathrm{NRV} = \frac{
        \sum_k w_k \bigl(s_k - \hat{s}_k\bigr)^2
    }{
        \sum_k w_k \bigl(s_k - \bar{s}_w\bigr)^2 + \epsilon
    },
    \label{eq:nrv}
\end{equation}
where $k$ index represents individual observations, $\bar{s}_w = \bigl(\sum_k w_k s_k\bigr) /
\bigl(\sum_k w_k\bigr)$ is the weighted mean signal, and
$\epsilon = 10^{-9}$~dBW$^2$ prevents division by zero. NRV near
zero indicates close conformance to the expected source geometry; NRV near
unity indicates an incoherent or diffuse pattern. The effective source size
is reported as $\mathrm{FWHM}_\mathrm{eff} = \sqrt{(\mathrm{FWHM}_x^2 +
\mathrm{FWHM}_y^2)/2}$, with $\mathrm{FWHM}_{x,y} =
2\sqrt{2\ln 2}\,\sigma_{x,y}$ from the $N=7$ free fit and held fixed for
all sub-constellation template fits.

\subsubsection{Persistence Monitoring and Activity Pattern Analysis}
\label{sec:persistence}

DDM noise floor measurements are assigned to discrete temporal windows of
duration $\Delta t = 90$~min, chosen to approximate the CYGNSS orbital period
of approximately 93~min. For a given geographic region $\mathcal{R}$ defined by
a latitude--longitude bounding box and a satellite $k$, each window $m$ is
assigned the mean noise floor over all observations falling within that window
and within $\mathcal{R}$:

\begin{equation}
    \bar{n}_{k}[m] =
    \begin{cases}
        \dfrac{1}{|\mathcal{S}_{k,m}|} \displaystyle\sum_{i \in \mathcal{S}_{k,m}} n_i
        & \text{if } |\mathcal{S}_{k,m}| \geq M_{\min}, \\[8pt]
        \texttt{NaN} & \text{otherwise,}
    \end{cases}
    \label{eq:window_mean}
\end{equation}

\noindent where $\mathcal{S}_{k,m}$ denotes the set of boresight-corrected DDM
observations from satellite $k$ inside $\mathcal{R}$ during window $m$, and
$M_{\min} = 5$ is the minimum number of valid DDM samples required to consider
a window populated. Windows with fewer than $M_{\min}$ samples are treated as
missing observations (\texttt{NaN}) to exclude geometrically marginal passes
that would otherwise introduce high measurement noise. In the fine-resolution
timeline figures, a window duration of 5~min with $M_{\min} = 1$ is used
exclusively for visualisation of the raw pass structure; all detection
computations use the 90-min grid.

For a sub-constellation of $N$ satellites drawn from the full seven-satellite
CYGNSS fleet $\mathcal{K} = \{\text{CYG01}, \ldots, \text{CYG08}\} \setminus
\{\text{CYG06}\}$, the merged noise floor series is formed as the element-wise
mean over the $N$ constituent satellite series:

\begin{equation}
    \bar{n}_{(N)}[m] = \frac{1}{|\{k : \bar{n}_k[m] \neq \texttt{NaN}\}|}
    \sum_{\substack{k \in \mathcal{C} \\ \bar{n}_k(m) \neq \texttt{NaN}}}
    \bar{n}_k[m],
    \label{eq:nanmean}
\end{equation}

\noindent where $\mathcal{C} \subseteq \mathcal{K}$ with $|\mathcal{C}| = N$.
This nanmean formulation assigns \texttt{NaN} to window $m$ only when all $N$
satellites in $\mathcal{C}$ lack a valid observation, preserving every window
for which at least one satellite has coverage. An alternative aggregation used
in the detection sensitivity analysis (Section~\ref{sec:persistence:detprob})
takes the element-wise maximum:

\begin{equation}
    \bar{n}_{(N)}^{\max}[m] =
    \max_{k \in \mathcal{C},\; \bar{n}_k[m] \neq \texttt{NaN}} \bar{n}_k[m],
    \label{eq:nanmax}
\end{equation}

\noindent which preserves the strongest signal observed by any member of the
sub-constellation in window $m$ and is discussed further below.

\begin{figure}[!b]
\centering
\includegraphics[width=\textwidth]{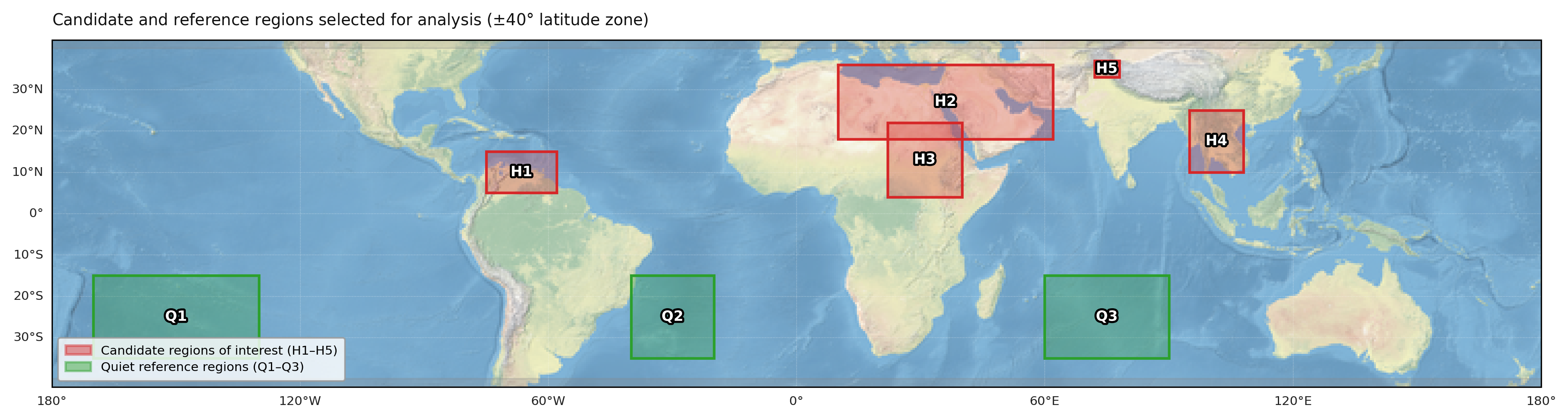}
\caption{Geographic locations of candidate regions of interest (H1--H5, red) and quiet reference regions (Q1--Q3, green) selected for analysis within the $\pm40\degree$ latitude zone. Region~H1: Venezuela, Region~H2: Middle East, Region~H3: Sudan/S.~Sudan, Region~H4: Myanmar/Laos/Cambodia, Region~H5: Kashmir. Region~Q1: S.~Pacific, Region~Q2: S.~Atlantic, Region~Q3: S.~Indian Ocean.}
\label{fig:region_map}
\end{figure}

Figure~\ref{fig:region_map} shows the geographic extent of the eight bounding boxes used in this study. The five candidate RFI hotspot regions and three quiet ocean reference regions are formally defined in Table~\ref{tab:region_definitions}. The five candidate regions were selected based on prior reporting of GNSS interference activity in the literature and open-source monitoring databases. The three quiet ocean regions serve as baseline references, providing an estimate of the ambient noise floor in the absence of terrestrial interference sources. The marked variation in bounding box area across candidate regions, from the relatively compact Region~H5 to the broad Region~H2, has direct implications for observational sampling density, as discussed in Section~\ref{results_persistence}.

\begin{table}[htbp]
\centering
\caption{Definition of candidate RFI hotspot regions (H1--H5) and quiet reference regions (Q1--Q3) used throughout this study. Identifiers are used in all subsequent results and discussion.}
\label{tab:region_definitions}
\setlength{\tabcolsep}{6pt}
\begin{tabular}{llp{10.0cm}}
\toprule
\textbf{ID} & \textbf{Region Name} & \textbf{Description / Selection Basis} \\
\midrule
H1 & Venezuela            & Episodic jamming activity; corroborated by open-source aviation notices \\
H2 & Middle East          & Sustained, high-intensity RFI; broad geographic extent \\
H3 & Sudan / S.\ Sudan    & Moderate RFI; Red Sea corridor activity \\
H4 & Myanmar / Laos / Cambodia & Multi-country interference corridor; Thailand--Myanmar border activity \\
H5 & Kashmir              & Compact mid-latitude target; geographically restricted bounding box \\
\midrule
Q1 & S.\ Pacific Ocean    & Quiet baseline; absence of terrestrial interference sources \\
Q2 & S.\ Atlantic Ocean   & Quiet baseline; absence of terrestrial interference sources \\
Q3 & S.\ Indian Ocean     & Quiet baseline; used as primary calibration reference \\
\bottomrule
\end{tabular}
\end{table}

At each window $m$, the $N{=}7$ merged series for all three quiet regions (Region~Q1--Region~Q3) are
pooled and collapsed to a single representative value via the element-wise
median. A rolling history of $H_w = 224$ windows (14~days) is then used to
compute, for each window $m$\footnote{The estimators are strictly causal: each depends only on the $H_w$
preceding samples $p[m-1],\ldots,p[m-H_w]$, following the running-estimator
convention of (\citet{proakis2007dsp}, Ch.~12. The index $m$ is used
(rather than $w$, reserved for window functions) to avoid notational collision.}:

\begin{align}
    \mu[m]    &= \operatorname{median}_{k=1}^{H_w}\, p[m - k],
    \label{eq:rolling_median} \\[4pt]
    \sigma[m] &= \operatorname{median}_{k=1}^{H_w}\, \bigl|p[m - k] - \mu[m]\bigr|,
    \label{eq:rolling_mad}
\end{align}

\noindent where $p[m]$ is the pooled quiet-region value at window $m$ and
$H_w = 224$ is the history length in windows. The estimator $\sigma[m]$ is the
median absolute deviation (MAD), which is robust to occasional elevated
measurements in the quiet regions and provides a scale estimate that does not
inflate in the presence of outliers. A minimum floor of
$\sigma_{\min} = 0.02$~dBW is applied to avoid division by near-zero values.
The first $H_w$ windows constitute a warm-up period during which the rolling
estimates lack sufficient history; all detections within this period are
suppressed.

For a hotspot region with merged series $\bar{n}_{(N)}[m]$, the detection
z-score at window $m$ is:

\begin{equation}
    z[m] = \frac{\bar{n}_{(N)}[m] - \mu[m]}{\sigma[m]},
    \label{eq:zscore}
\end{equation}

\noindent where $z[m]$ is $\texttt{NaN}$ when $\bar{n}_{(N)}[m]$ is $\texttt{NaN}$ or
$m$ falls within the warm-up period. A window is flagged as anomalous when
$z[m] > \zeta = 3.0$. To reduce false alarms from isolated noisy passes, a
confirmed detection requires a streak of at least $L = 2$ consecutive flagged
windows:

\begin{equation}
    \text{confirmed}[m] = 1 \iff z[m] > \zeta
    \;\text{ and }\; z[m+1] > \zeta,
    \label{eq:streak}
\end{equation}

\noindent with both windows non-$\texttt{NaN}$. The streak criterion is applied only to the
nanmean aggregation. For the nanmax aggregation (Eq.~\ref{eq:nanmax}), which
already represents the most sensitive available observation at each window, a
single window above threshold is considered sufficient evidence, and the streak
requirement is relaxed to $L = 1$.

This asymmetry is physically motivated: under nanmean, the merged series
averages contributions from multiple satellites including those observing
background conditions, attenuating the RFI signature and requiring sustained
elevation for confirmation. Under nanmax, the merged value reflects the
strongest single-satellite observation in the window, which is an unambiguous
upper bound on what any member of the constellation saw; a single exceedance
above $3\sigma$ is therefore a stronger statement about the local environment.

All detection timing and reliability statistics are computed over the full
combinatorial ensemble of sub-constellations. For constellation size $N$, there
are $\binom{7}{N}$ distinct sub-constellations: 7 for $N{=}1$, 35 for $N{=}3$,
and 1 for $N{=}7$. For each combination $\mathcal{C}$, the detector is run
independently on the merged series derived from exactly those $N$ satellites,
and the index $m^*(\mathcal{C})$ of the first confirmed detection window is
recorded ($\texttt{NaN}$ if none recorded). The reported first-detection date for
constellation size $N$ is the median of $m^*(\mathcal{C})$ over all
combinations that produce a detection, expressed as a calendar date.

The detection probability curve at day $d$ is:

\begin{equation}
    P_{\mathrm{det}}(N, d) = \frac{1}{\binom{7}{N}}
    \sum_{\mathcal{C}} \mathbf{1}\!\left[
        \lfloor m^*(\mathcal{C}) / W_d \rfloor \leq d
    \right],
    \label{eq:pdet}
\end{equation}

\noindent where $W_d = 16$ is the number of 90-min windows per day.
The term $\mathbf{1}[\cdot]$ is the indicator function, returning 1 if the
condition inside is true and 0 otherwise; it is set to zero for any
combination $\mathcal{C}$ that produces no confirmed detection (i.e.\
$m^*(\mathcal{C}) = \texttt{NaN}$). Concretely, the floor
$\lfloor m^*(\mathcal{C})/W_d \rfloor$ converts the window index of first
detection into a day number, and the sum counts how many of the
$\binom{7}{N}$ sub-constellations have confirmed a detection by day $d$.
Dividing by $\binom{7}{N}$ yields an empirical probability between 0 and 1.
This formulation captures both the speed of detection (rate of rise of the
curve) and its reliability (asymptotic value), without mixing the two
into a single point statistic.


\begin{figure}[!t]
\centering
\begin{minipage}{0.49\textwidth}
\centering
\includegraphics[width=\linewidth]{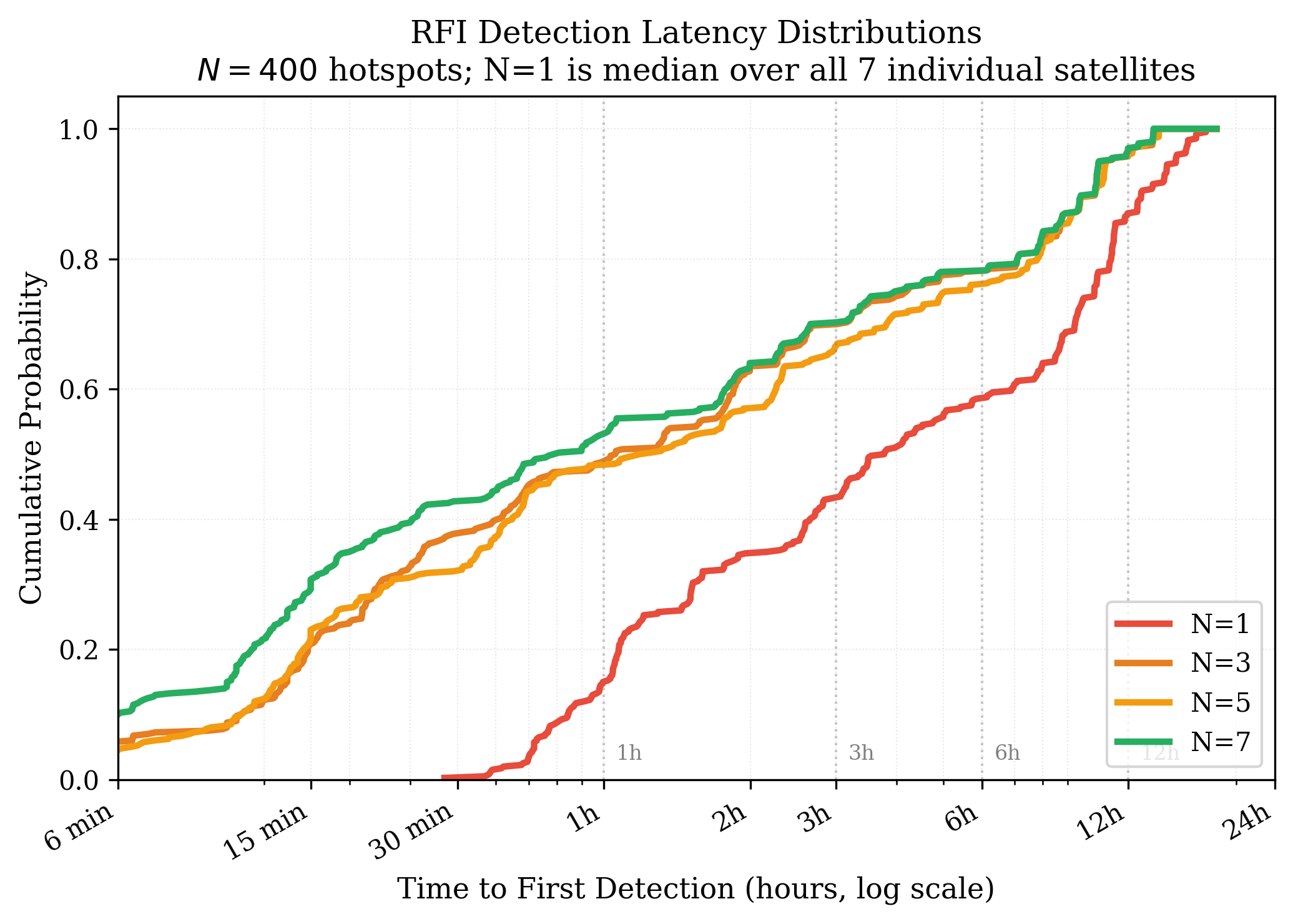}
\caption{CDF of detection latency for persistent RFI hotspots from observation start (log scale). $N=1$ is the median across satellites; $N=3$ and $N=5$ are medians across sub-constellations.}
\label{fig:detection_cdf}
\end{minipage}
\hfill
\begin{minipage}{0.49\textwidth}
\centering
\includegraphics[width=\linewidth]{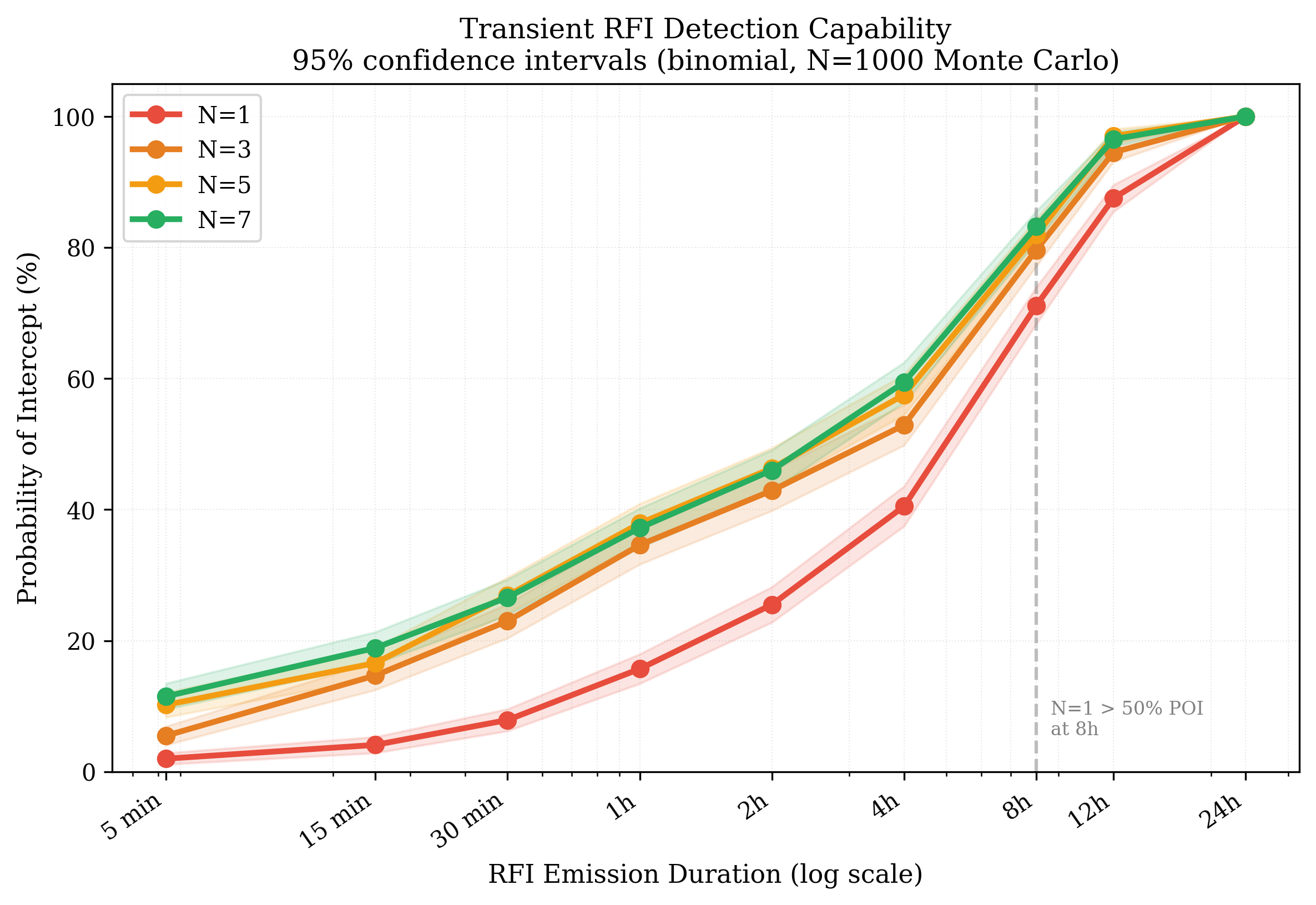}
\caption{Probability of intercept for transient RFI emissions of varying duration (5 min–24 h, log scale) from 1000 Monte Carlo trials. Shaded regions denote 95\% confidence intervals.}
\label{fig:transient_poi}
\end{minipage}
\end{figure}

\section{Results}
\label{sec:results}

This section evaluates how constellation size influences the capability of CYGNSS GNSS-R observations to detect, localize, and monitor persistent and transient RFI sources. The analysis examines four complementary dimensions of performance: temporal coverage, spatial coverage, geometric coherence, and persistence monitoring. Comparisons are performed across sub-constellation configurations ranging from a single satellite ($N{=}1$) to the full seven-satellite constellation ($N{=}7$).

\subsection{Temporal Coverage Analysis}

Figure~\ref{fig:detection_cdf} shows the cumulative distribution of
time-to-first-detection across 400 high noise hotspots. The full
constellation ($N=7$) achieves a median detection latency of 48~minutes
compared to 3~hours 46~minutes for individual satellites ($N=1$),
representing a $4.7\times$ improvement. The constellation advantage is
most pronounced in the first few hours: $N=7$ detects 53\% of sources
within one hour compared to only 15\% for $N=1$. By 24~hours, all
configurations achieve 100\% detection, but the early-detection advantage
is operationally critical.

Table~\ref{tab:threshold_detection} presents detection rates within
operationally relevant time windows. Within three hours, $N=7$ detects
70\% of sources compared to 43\% for $N=1$. The 90th-percentile latencies
differ by less than one minute across $N=3$, $N=5$, and $N=7$
configurations (approximately 10~hours 14--15~minutes), suggesting
diminishing returns for the most challenging locations where geographic
accessibility dominates over constellation density. The mean standard
deviation of 1~hour 58~minutes in $N=1$ detection times across hotspots
quantifies the sensitivity to satellite orbital phase at analysis start.

\begin{table}[htbp]
\centering
\caption{Fraction of RFI hotspots detected within time threshold by
constellation size ($N=400$ hotspots).}
\label{tab:threshold_detection}
\begin{tabular}{lcccc}
\toprule
Window length & $N=1$ & $N=3$ & $N=5$ & $N=7$ \\
\midrule
1 hour   & 15.0\% & 48.8\% & 48.2\% & 53.0\% \\
3 hours  & 43.2\% & 69.8\% & 66.5\% & 70.0\% \\
6 hours  & 58.5\% & 78.0\% & 76.0\% & 78.0\% \\
12 hours & 87.0\% & 96.8\% & 96.0\% & 97.0\% \\
24 hours & 100.0\% & 100.0\% & 100.0\% & 100.0\% \\
\bottomrule
\end{tabular}
\end{table}

Figure~\ref{fig:transient_poi} quantifies the probability of intercepting
brief RFI emissions through 1,000-scenario Monte Carlo simulation per
duration. Shaded bands represent 95\% binomial confidence intervals. The
constellation advantage scales inversely with emission duration: for
5-minute bursts, $N=7$ achieves 11.5\% POI compared to 2.0\% for $N=1$,
a 5.8 times improvement. For one-hour emissions, $N=7$ (37.2\%)
outperforms $N=1$ (15.7\%) by 2.4 times. At eight hours, $N=1$ reaches
71.1\% POI and the constellation advantage diminishes to 1.2 times. By
24~hours, all configurations achieve 100\% POI as multiple orbital cycles
guarantee coverage. 
\begin{table}[!t]
\centering
\caption{Revisit time statistics for the top 50 RFI hotspots by noise
intensity over a 19-day observation window (December~10--28, 2025).}
\label{tab:revisit_time}
\begin{tabular}{lcccc}
\toprule
Config & Median & Mean & 90th Percentile & Minimum \\
\midrule
$N=1$ & 23h 09min & 27h 37min & 61h 16min & 1h 36min \\
$N=3$ &  8h 14min & 10h 23min & 23h 10min &   15 min \\
$N=5$ &  2h 45min &  6h 39min & 15h 39min &   15 min \\
$N=7$ &  1h 49min &  5h 19min & 14h 06min &   15 min \\
\bottomrule
\end{tabular}
\end{table}

\begin{figure}[!t]
\centering
\includegraphics[width=0.90\textwidth]{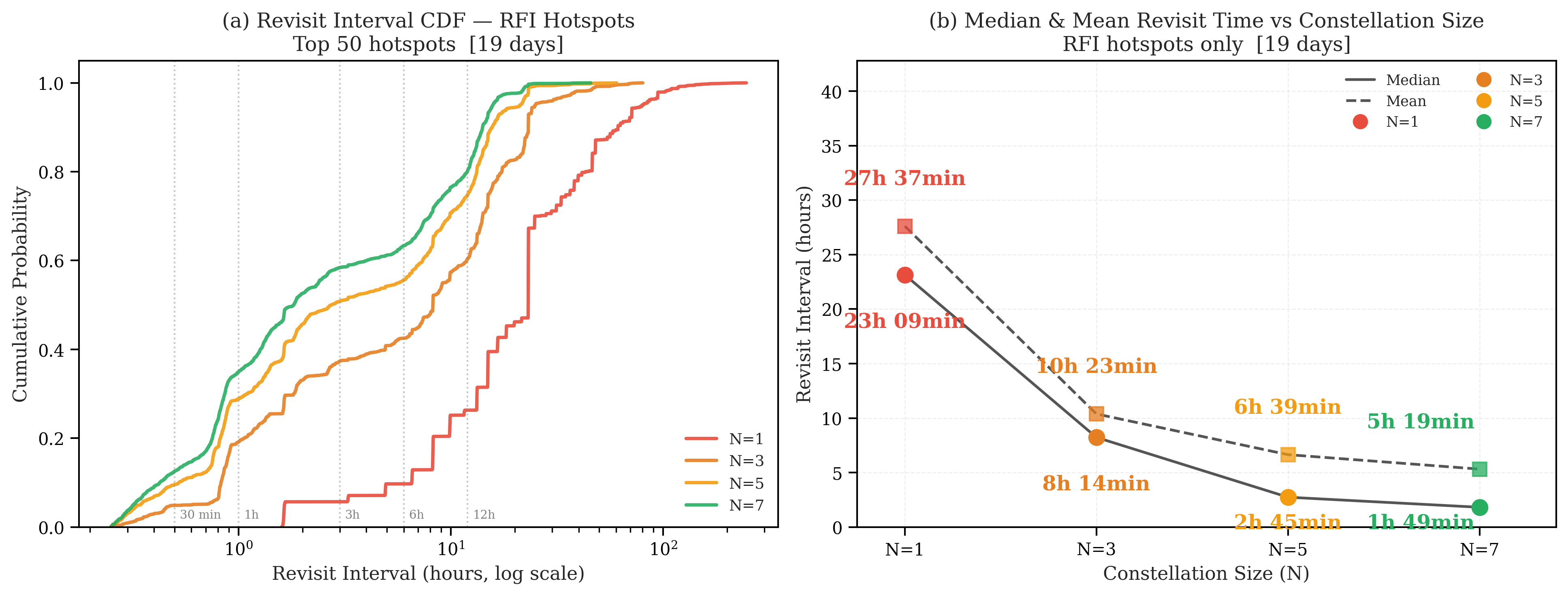}
\caption{Revisit statistics for the top 50 RFI hotspots over a 19-day window: (a) CDF of inter-pass intervals (log scale); (b) median (solid) and mean (dashed) revisit intervals versus constellation size.}
\label{fig:revisit_time}
\end{figure}

The crossover analysis indicates that for emissions
shorter than two hours, constellation deployment provides greater than 2 times POI improvement, demonstrating that multi-satellite systems are
particularly advantageous for detecting brief interference events.
Figure~\ref{fig:detection_cdf} and Figure~\ref{fig:transient_poi},
displayed side by side, highlight that real high-noise hotspots and
simulated brief RFI emissions produce very similar detection patterns: in
both cases, multi-satellite constellations significantly improve the
probability of interception for short events, confirming that the
simulations realistically reflect the observed challenges of transient
interference.

Figure~\ref{fig:revisit_time}(a) shows the cumulative distribution of
inter-pass gaps for the top 50 RFI hotspots over a 19-day observation
window. Individual satellites ($N=1$) exhibit a median revisit interval
of 23~hours 9~minutes, reflecting the relatively sparse ground track
density of a single CYGNSS satellite over a fixed point. Expanding to
$N=3$ reduces the median to 8~hours 14~minutes, while $N=5$ and $N=7$
further compress it to 2~hours 45~minutes and 1~hour 49~minutes
respectively. Figure~\ref{fig:revisit_time}(b) illustrates this
monotonic improvement alongside mean revisit times, which are
substantially higher due to the long-tail distribution of inter-pass
gaps: mean values reach 27~hours 37~minutes for $N=1$ and 5~hours
19~minutes for $N=7$. Table~\ref{tab:revisit_time} summarises revisit
statistics across configurations. The 90th-percentile gap decreases from
61~hours 16~minutes for $N=1$ to 14~hours 6~minutes for $N=7$,
representing a 4.3 times improvement in worst-case monitoring latency.
These results indicate that while the full constellation meaningfully
reduces typical revisit intervals, the long tail of the distribution
reflects periods of unfavourable orbital geometry and highlights the
persistent coverage challenge inherent to low-Earth orbit constellations
of this size.

\subsection{Spatial Coverage Analysis}

\begin{figure}[!b]
\centering
\includegraphics[width=0.9\textwidth]{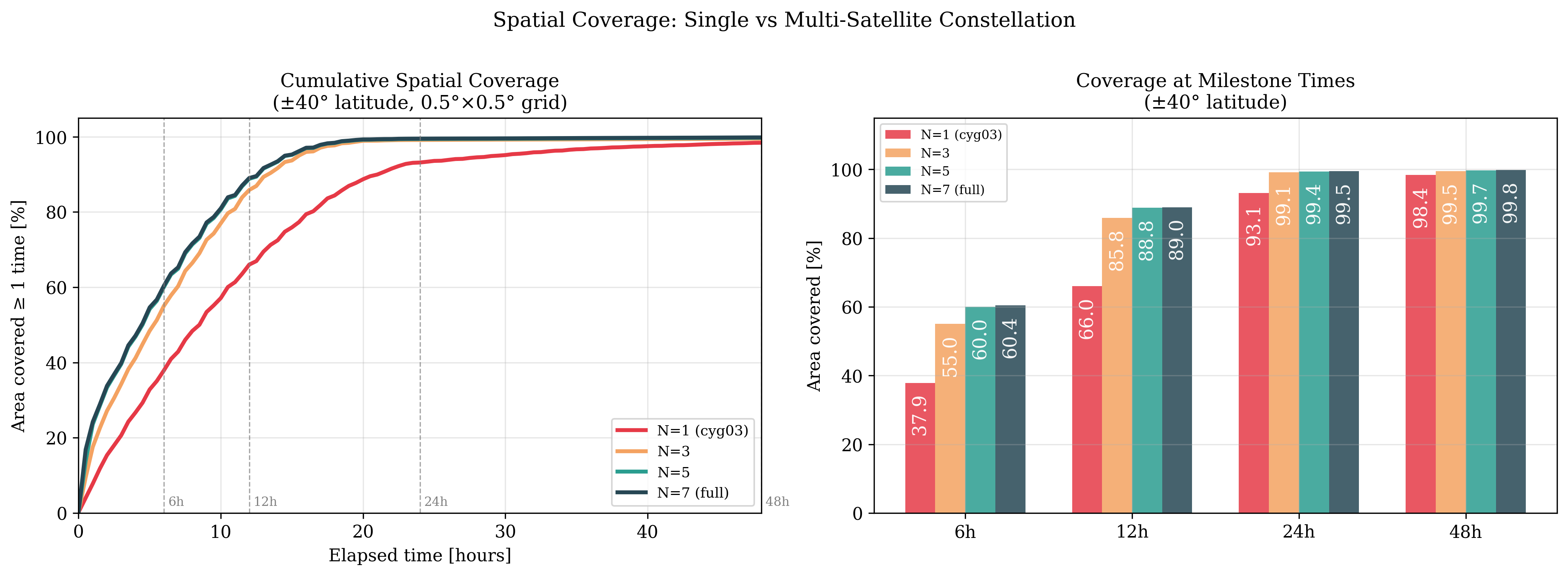}
\caption{Cumulative spatial coverage fraction over 48 hours (left) and coverage at milestone times (right) for constellation configurations N=1, N=3, N=5, and N=7.}
\label{fig:spatial_coverage}
\end{figure}

\begin{table}[htbp]
\centering
\caption{Spatial coverage fraction at milestone times for different constellation configurations over a 48-hour period (Dec~17--18, 2025).}
\label{tab:spatial_coverage}
\begin{tabular}{lcccc}
\toprule
Configuration & 6h & 12h & 24h & 48h \\
\midrule
N=1 (cyg03) & 37.9\% & 66.0\% & 93.1\% & 98.4\% \\
N=3 & 55.0\% & 85.8\% & 99.1\% & 99.5\% \\
N=5 & 60.0\% & 88.8\% & 99.4\% & 99.7\% \\
N=7 (full) & 60.4\% & 89.0\% & 99.5\% & 99.8\% \\
\bottomrule
\end{tabular}
\end{table}

\begin{figure}[htbp]
\centering
\includegraphics[width=\textwidth]{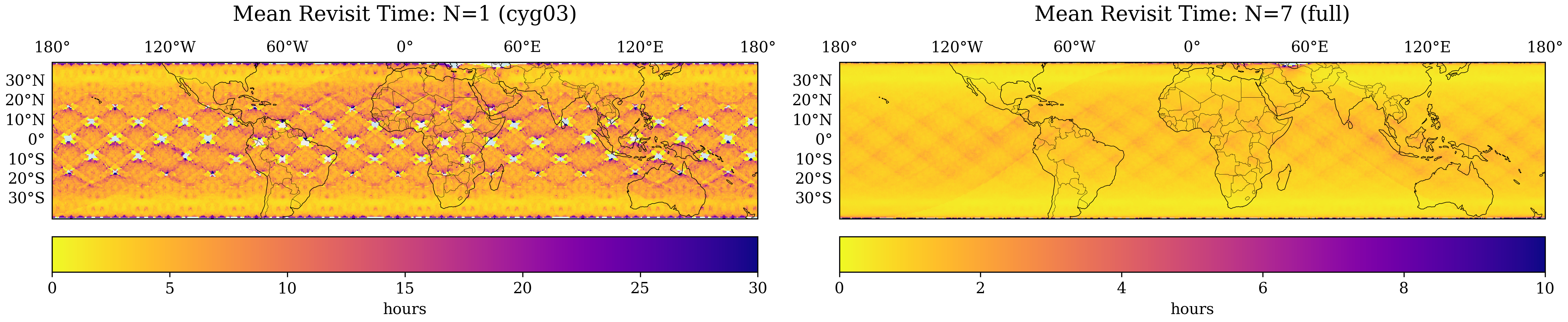}
\caption{Geographic distribution of mean GNSS-R footprint revisit time over 48 hours for single-satellite ($N=1$, left) and full constellation ($N=7$, right) configurations. The color scale indicates mean time between consecutive satellite passes in hours.}
\label{fig:revisit_mean}
\end{figure}

\begin{figure}[htbp]
\centering
\includegraphics[width=\textwidth]{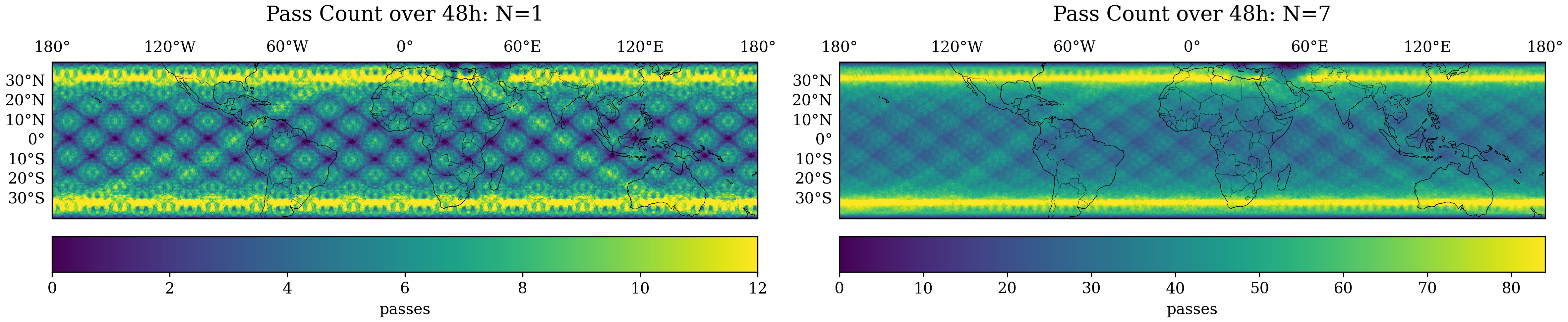}
\caption{Total number of GNSS-R footprint passes per $0.5^\circ \times 0.5^\circ$ grid cell over 48 hours for single-satellite ($N=1$, left) and full constellation ($N=7$, right) configurations.}
\label{fig:revisit_passes}
\end{figure}

Figure~\ref{fig:spatial_coverage} shows the cumulative spatial coverage and milestone-based comparison across constellation sizes. The N=7 constellation achieves 60.4\% coverage within 6 hours, increasing to 99.8\% by 48 hours. In contrast, the single-satellite baseline (N=1, CYGNSS FM03) reaches only 37.9\% at 6 hours and 98.4\% at 48 hours. Table~\ref{tab:spatial_coverage} presents coverage fractions at key time milestones. The marginal improvement from N=5 to N=7 reflects the high baseline coverage achieved with five satellites; the primary benefit of additional satellites lies in reducing revisit gaps rather than expanding total coverage area.

Figure~\ref{fig:revisit_mean} displays the mean revisit time across the $\pm 40^\circ$ latitude zone for both single-satellite ($N=1$) and full constellation ($N=7$) configurations. The single-satellite configuration exhibits strong geographic variability, with mean revisit times ranging from under 5 hours near the equator to over 25 hours at high latitudes approaching $\pm 35^\circ$. This latitudinal dependence arises from CYGNSS's $35^\circ$ orbital inclination: regions near the orbit's turnpoints experience sparser coverage due to slower ground-track progression. The $N=7$ constellation dramatically reduces both the mean revisit time and its spatial variability. Median revisit time improves from 5.8 hours ($N=1$) to 1.0 hour ($N=7$), representing a 5.8$\times$ enhancement in temporal sampling density. The 90th percentile revisit time decreases from 10.5 hours to 1.5 hours, indicating that even poorly-sampled regions benefit substantially from constellation density. Over the 48-hour analysis window, 1,846 cells (1.6\% of the grid) were never visited by the single satellite, compared to only 214 cells (0.2\%) for the full constellation.

Figure~\ref{fig:revisit_passes} shows the total number of footprint passes per grid cell over the 48-hour window. The single satellite achieves a median of approximately 8 passes per cell, while the full constellation provides approximately 49 passes. This demonstrates the multiplicative effect of constellation size on temporal sampling. The geographic patterns in both figures reveal persistent coverage asymmetries: equatorial regions ($\pm 10^\circ$ latitude) receive more frequent revisit due to denser orbital cross-track spacing, while near-turnpoint latitudes ($30^\circ$--$40^\circ$) exhibit sparser coverage. This latitudinal gradient is intrinsic to the orbital geometry and persists across all constellation sizes, though its magnitude diminishes as $N$ increases.

The checkerboard pattern of unvisited cells visible in the $N=1$ map (white patches, distinct from the color-scale minimum) reflects the discrete, swath-limited nature of CYGNSS specular point geometry. GNSS-R specular points are tied to the instantaneous geometry between GPS transmitters, the Earth's surface, and the receiver. A single satellite tracks a sparse, structured set of ground tracks that within a 48-hour window leave persistent inter-swath gaps. These gaps produce a quasi-periodic spatial pattern whose scale is set by the orbital period and Earth's rotation rate. The $N=7$ constellation as seen on Figure~\ref{fig:revisit_mean} fills these gaps through spatial diversity: the seven satellites occupy different orbital planes and phases, ensuring that inter-swath voids of one satellite are covered by the ground tracks of others.

These results quantify the fundamental trade-off between spatial coverage and temporal resolution in space-based GNSS-R RFI monitoring. A single satellite provides broad geographic reach (98.4\% of $\pm 40^\circ$ zone within 48 hours) but sparse temporal sampling (median 5.8-hour revisit), making it poorly suited for detecting brief or intermittent interference events. The full seven satellite constellation maintains near-complete spatial coverage (99.8\%) while achieving hourly revisit times, enabling detection of transient RFI sources that would be missed by single-satellite sparse sampling. Combined with the temporal coverage improvements demonstrated in Section~\ref{sec:temporal}, the full constellation provides both rapid initial detection (median 2.8-hour latency from emission onset) and sustained monitoring (hourly re-observation) across the global $\pm 40^\circ$ latitude zone.

\subsection{Geometric Coverage and Spatial Coherence}
\label{sec:results_geometric}

\begin{figure}[htbp]
    \centering
    \includegraphics[width=0.90\linewidth]{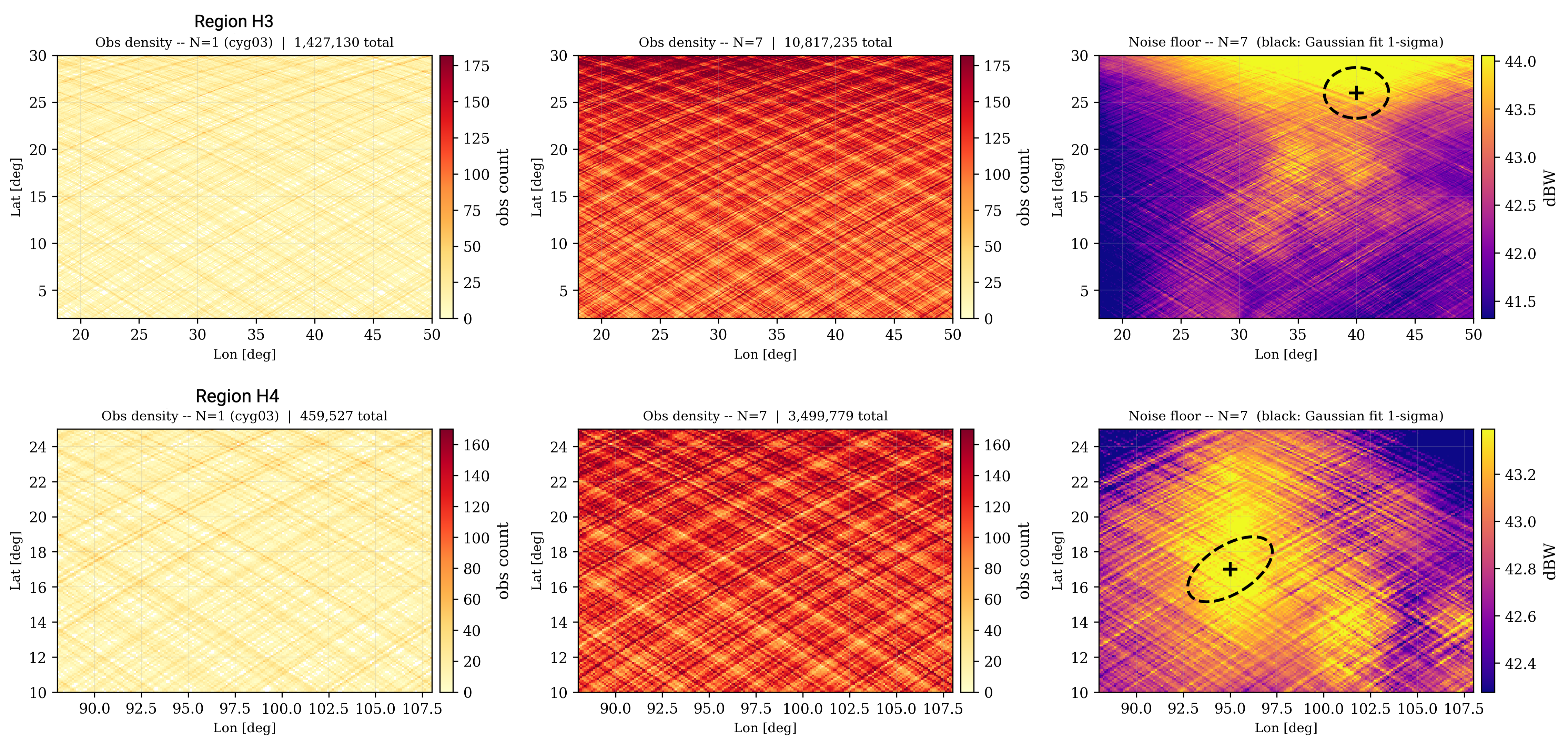}
    \caption{Spatial maps for Regions H3 (top) and H4 (bottom) over October--December 2025. Left: observation density for $N=1$. Center: observation density for $N=7$. Right: mean noise-floor elevation relative to the quiet-ocean background with fitted 1-$\sigma$ Gaussian ellipse and source center.}
    \label{fig:gc_spatial_maps}
\end{figure}

Spatial coherence analysis was performed for Region~H3 and Region~H4 over the
92-day period from 1~October to 31~December 2025. Both regions exhibited mean
noise floor elevations above the quiet ocean reference of $+0.773$~dBW and
$+1.210$~dBW respectively under $N=7$, satisfying the minimum elevation
threshold of $0.30$~dBW across all constellation configurations.

Figure~\ref{fig:gc_spatial_maps} establishes the spatial sampling context.
The left column shows the observation density accumulated by the single-satellite
baseline ($N=1$, cyg03), and the centre column shows the corresponding $N=7$
density. For Region~H3, $N=1$ accumulates 1,427,130 observations against
10,817,235 for $N=7$, by a factor of 7.6 times; for Region~H4 the ratio is
identical at 459,527 versus 3,499,779. In both $N=1$ density maps, individual
orbital ground tracks are visible as diagonal stripes with inter-swath gaps of
several hundred kilometres. The $N=7$ maps fill these gaps nearly completely,
producing spatially uniform coverage across both domains. The right column shows
the $N=7$ mean noise floor elevation with the fitted Gaussian ellipse overlaid.
For Region~H4, the ellipse is centred near
$(95^\circ\text{E},\,17^\circ\text{N})$, coinciding with known interference
activity along the Thailand--Myanmar border region. For Region~H3, the ellipse
sits over the elevated northeastern portion of the domain, consistent with
concentrated activity in the Red Sea corridor. In both cases the noise floor
gradient, rising toward the ellipse centre and decaying at the periphery, is
visually consistent with the Gaussian source model.

\begin{figure}[!t]
    \centering
    \includegraphics[width=0.81\linewidth]{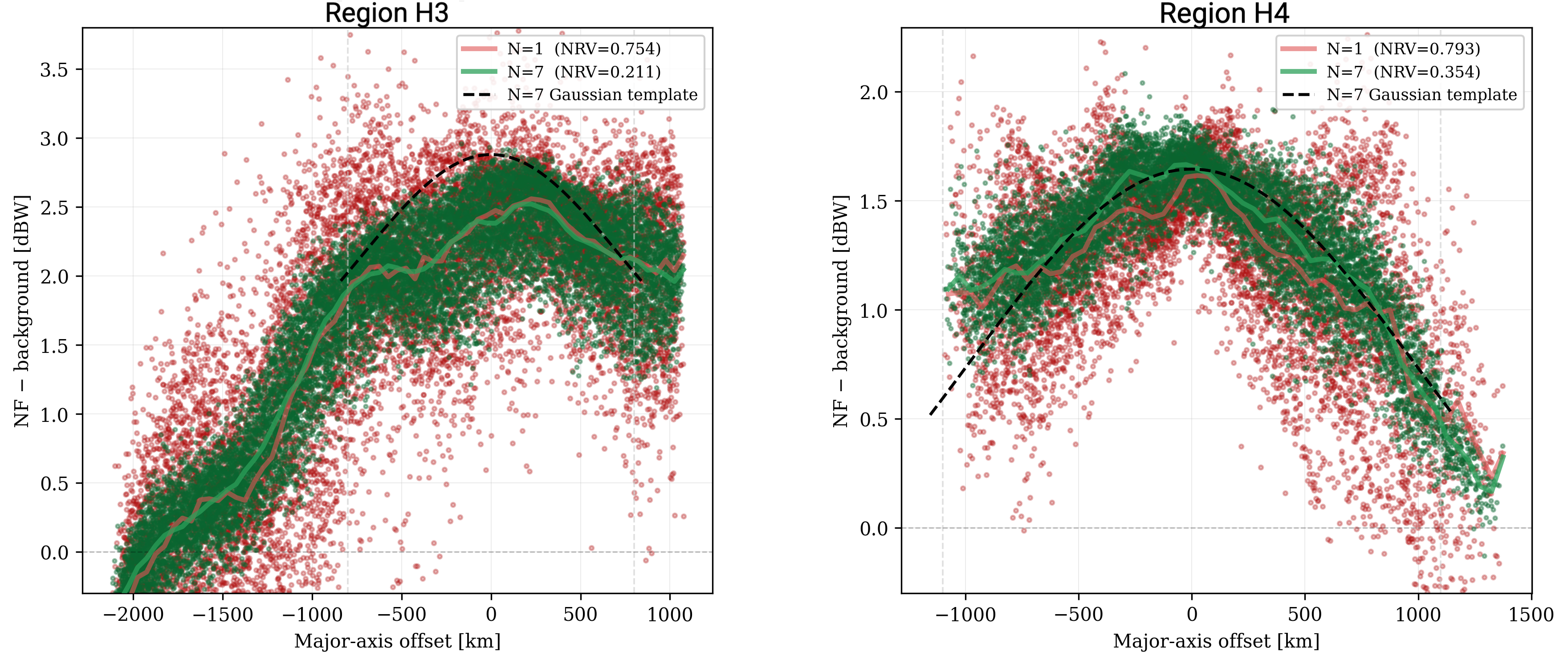}
    \caption{Major-axis cross-sections for Regions H3 (left) and H4 (right). Points denote $0.1^\circ$ grid cells, solid lines show 50-km bin medians, and dashed curves indicate the $N=7$ Gaussian template. Vertical markers denote fit-radius boundaries.}
    \label{fig:gc_profiles_overlay}
\end{figure}

Figure~\ref{fig:gc_profiles_overlay} shows the one-dimensional cross-section
profiles along the Gaussian major axis for both hotspots, with $N=1$ and $N=7$
superimposed on a common axis. Each point is a single $0.1^\circ$ grid cell
projected onto the major-axis direction; the solid lines are 50-km bin medians;
the dashed curve is the $N=7$ Gaussian template. For Region~H3, the $N=7$ bin
median closely tracks the template across the full $\pm 2000$~km cross-section
extent, rising smoothly to a peak near the origin and decaying symmetrically on
both sides. The $N=1$ bin median follows the same broad trend but with markedly
greater scatter: individual cell values span a $\pm 1$~dBW band around the
median, and the median itself oscillates at the scale of individual orbital
swath widths. For Region~H4, the contrast is equally pronounced; the $N=7$
median defines a clean, well-resolved peak near zero offset with a gradual decay
toward the fit radius boundary, while the $N=1$ median is noisier throughout
and the cloud of individual points is visibly more dispersed, particularly on
the negative-offset flank where single-satellite coverage is sparser. Both
panels show that the $N=7$ data points form a substantially denser and more
compact envelope around the median line than $N=1$, a direct consequence of
the 7.6 times increase in spatial sampling density.

\begin{figure}[!t]
    \centering
    \includegraphics[width=0.55\linewidth]{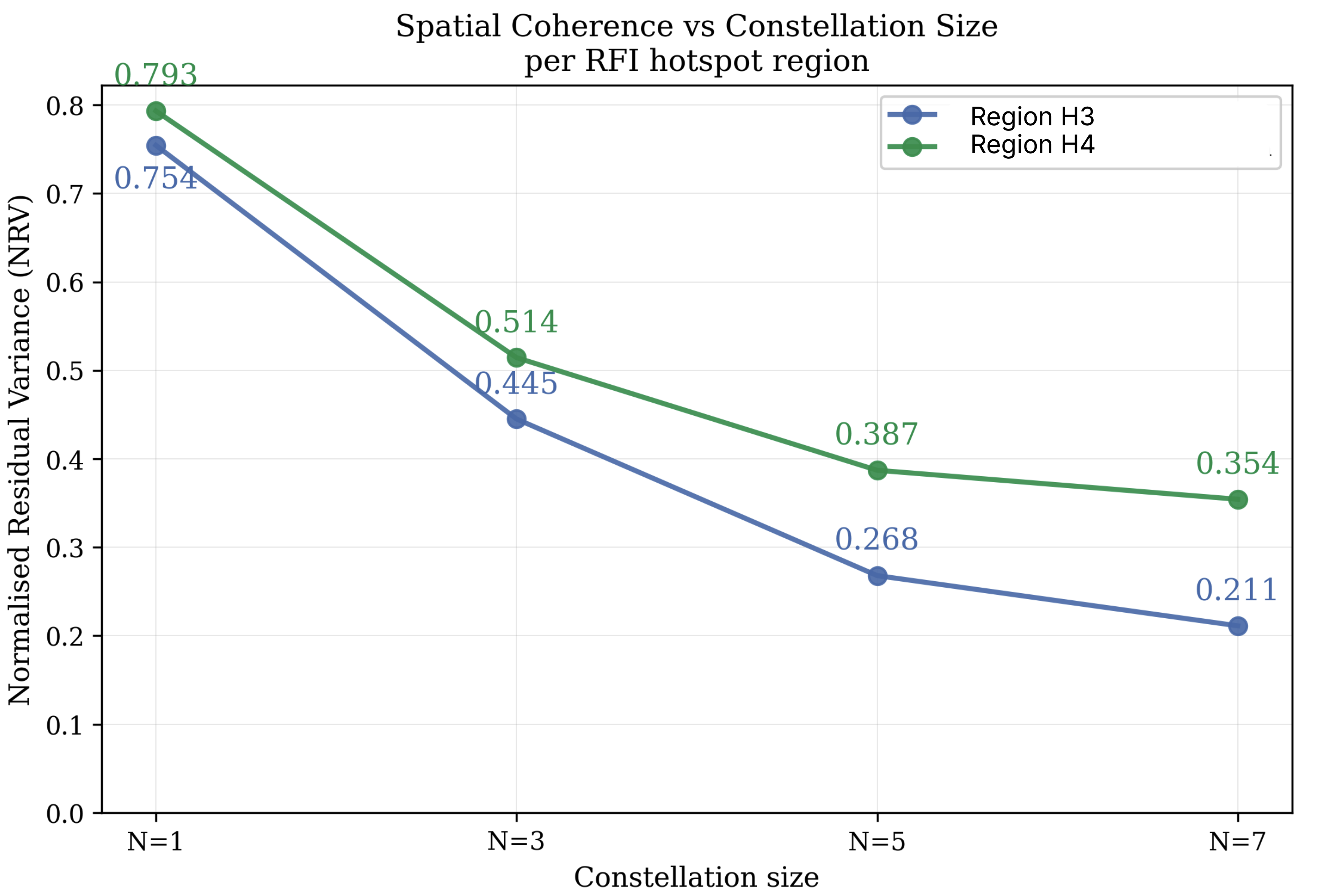}
    \caption{Normalised residual variance (NRV) as a function of number of satellites for Region~H3 and Region~H4. Lower NRV indicates closer conformance
    of the observed noise floor distribution to the $N=7$ Gaussian source
    template.}
    \label{fig:gc_nrv_vs_N}
\end{figure}

The NRV results are summarised in Figure~\ref{fig:gc_nrv_vs_N}.
For Region~H3, NRV decreases monotonically from 0.754 at $N=1$ to 0.211 at
$N=7$, a reduction of 72.0\%. The steepest improvement occurs between $N=1$
and $N=3$ (0.754 to 0.445), with continued gains through $N=5$ (0.268) and
$N=7$ (0.211). For Region~H4, the NRV follows the same monotonic trend from
0.793 at $N=1$ to 0.354 at $N=7$, a reduction of 55.3\%. The absolute NRV
values for Region~H4 are consistently higher than for Region~H3 at every
constellation size, which is consistent with the broader and more geographically
dispersed character of that interference region. The fitted effective FWHM of
2463~km for Region~H4 compared to 1423~km for Region~H3 reflects a source
distribution that spans three countries and produces a shallower, less sharply
peaked spatial gradient, a distribution that is intrinsically harder to
characterise with a single Gaussian model regardless of observation density.
Nonetheless, the monotonic NRV reduction with increasing $N$ holds for both
hotspots, confirming that the constellation-size effect is a property of the
spatial sampling geometry rather than a feature of any particular source.

Because the $N < 7$ fits use a fixed spatial template derived from the full
constellation, NRV measures a single well-defined quantity: how consistently
the sparser constellation's observations conform to the spatial structure that
the $N=7$ map has established. With $N=1$, the orbital ground track pattern
leaves large unsampled voids within the fit disc, and the observations that do
exist are concentrated in diagonal stripes aligned with the satellite's ground
track direction. When these anisotropically distributed samples are tested
against an isotropic Gaussian template, the residuals are necessarily large,
not because the source geometry has changed, but because the irregular sampling
cannot confirm it. As $N$ increases, the inter-track voids fill progressively,
the angular distribution of observations becomes more isotropic, and the
residuals decrease accordingly. The 72\% NRV reduction achieved for Region~H3
from $N=1$ to $N=7$ directly quantifies the fraction of geometric information
recoverable as a function of constellation size: a single satellite leaves
approximately three quarters of the source's spatial coherence unresolved,
while the full constellation reduces the unexplained spatial variance to
roughly one fifth of its single-satellite value.

\begin{figure}[!b]
    \centering
    \includegraphics[width=0.80\linewidth,
                     height=0.6\textheight,
                     keepaspectratio]{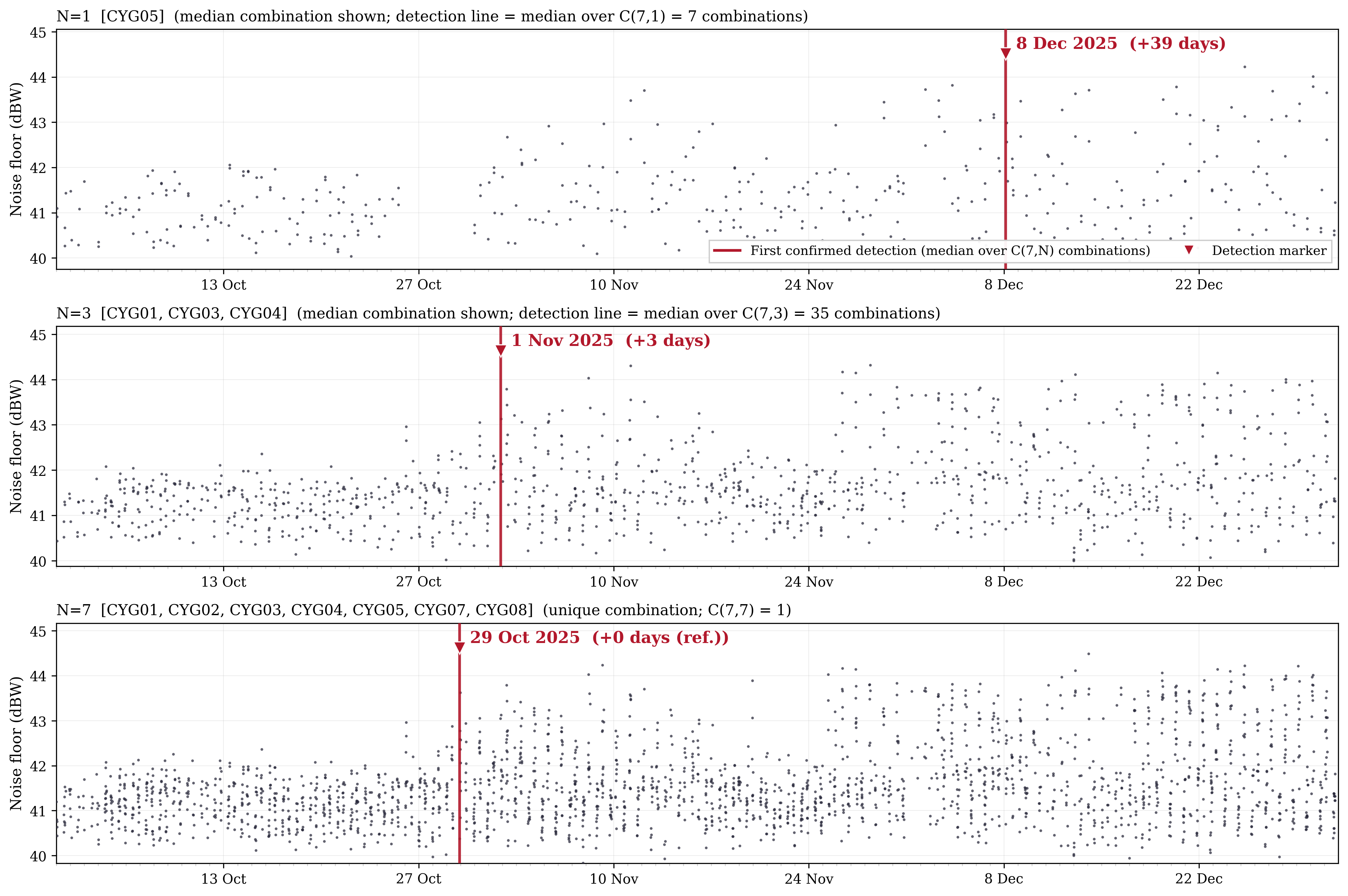}
    \caption{Region~H1 observation timeline at 5-min resolution for the
    best-coverage sub-constellation of each size.
    Each dot denotes a populated 5-min window satisfying the minimum observation
    threshold. The vertical red line marks the median first confirmed detection
    date across all $\binom{7}{N}$ sub-constellation combinations.}
    \label{fig:raw_timeline}
\end{figure}

\subsection{Persistence Monitoring and Activity Pattern Analysis}
\label{results_persistence}

\subsubsection{Observation Timeline and Coverage Structure}

Figure~\ref{fig:raw_timeline} presents the raw observation timeline over
Region~H1 for the best-coverage sub-constellation of each size, computed at a
5-min temporal resolution with a minimum of one DDM observation per window. Each
dot represents a single populated 5-min window; gaps correspond to periods
during which no satellite in the sub-constellation had sufficient coverage over
the region.

The figure demonstrates the central claim of multi-satellite persistence
monitoring. At $N{=}1$, CYG01 produces isolated clusters of observations
separated by multi-hour gaps; the median inter-observation gap is on the order
of several hours. At $N{=}7$, the combined fleet achieves near-continuous
coverage, with substantially fewer and shorter gaps. This difference in coverage
density has a direct and quantifiable consequence for detection timing: the
first confirmed detection under the nanmean detector occurs on 29~October~2025
for $N{=}7$, versus 1~November~2025 for $N{=}3$ and 8~December~2025 for
$N{=}1$: a 39-day advantage of the full constellation over a single satellite.

The mechanism linking coverage density to detection latency is the streak
confirmation requirement (Eq.~\ref{eq:streak}). For a single satellite
observing Region~H1 at a typical rate of 2--3 passes per day, each pass
populates one or two 90-min windows before the satellite moves out of the
bounding box. The next populated window is typically 6--8~hours later,
spanning several empty slots. Consequently, even when a given pass yields a
z-score above $\zeta = 3.0$, the immediately following 90-min window is
\texttt{NaN} and the streak resets to zero. Genuine RFI signatures observed by a
single satellite are therefore systematically filtered out by the streak
criterion whenever the orbital geometry does not happen to produce two
successive passes within 90~min. For $N{=}7$, consecutive 90-min slots are
populated by different satellites on independent orbits, satisfying the streak
criterion as soon as the regional noise floor rises above threshold, without
depending on any single satellite's phasing.

\subsubsection{Activity Pattern Heatmap}
\begin{figure}[htbp]
    \centering
    \includegraphics[width=0.85\linewidth]{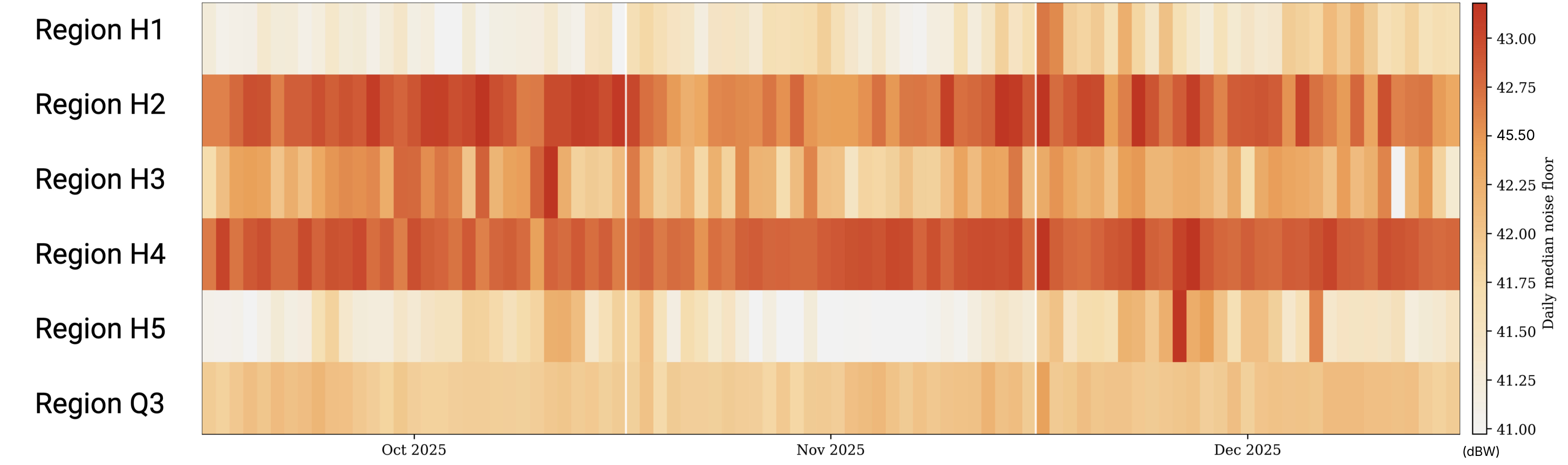}
    \caption{Daily median z-score for October 2025 to December 2025 relative to the quiet rolling baseline for
    Regions H1--H5 and quiet reference Region~Q3, computed using the $N{=}7$
    merged series.}
    \label{fig:heatmap}
\end{figure}

Figure~\ref{fig:heatmap} presents the daily median noise floor (dBW) over the
three-month analysis period for all five RFI hotspot regions alongside a quiet
ocean reference. Region~Q3 serves as a calibration baseline, reflecting the
ambient noise floor in the absence of terrestrial interference sources.
Region~H4 and Region~H2 exhibit the most sustained and intense elevated noise
floors throughout the period, remaining consistently above threshold.
Region~H1 displays a clearly intermittent pattern, consistent with episodic
jamming activity that activates and subsides across the period. Region~H3 and
Region~H5 show moderate, spatially variable elevation with occasional quiet
intervals, suggesting either lower-intensity or geographically localized sources
that fall in and out of the CYGNSS footprint.

Vertical white columns indicate days with no CYGNSS observations over a given
region rather than true signal absence, a consequence of both the sparse revisit
sampling discussed in Section~\ref{sec:spatial} and the varying geographic
extent of each region as seen in Figure~\ref{fig:region_map}. Smaller regions
such as Region~H5 encompass fewer grid cells, making them more susceptible to
complete observational gaps on any given day compared to broader regions like
Region~H2 or Region~H4.

\subsubsection{Detection Probability as a Function of Constellation Size}
\label{sec:persistence:detprob}

\begin{figure}[!b]
    \centering
    \includegraphics[width=0.83\linewidth, keepaspectratio]{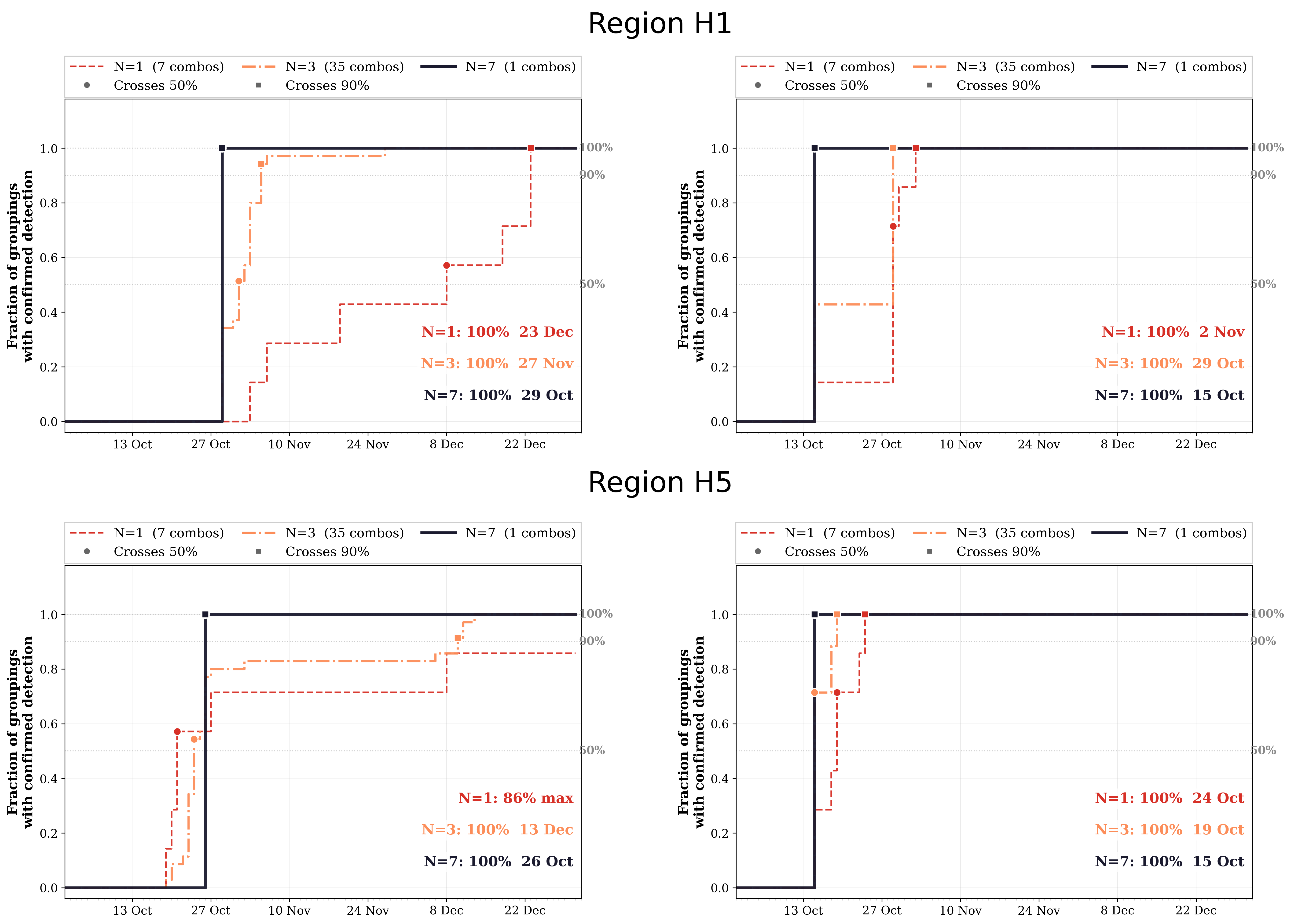}
    \caption{Detection probability versus time for Regions H1 and H5 under nanmean aggregation with streak confirmation ($L=2$, left) and nanmax aggregation ($L=1$, right). Curves show the fraction of sub-constellations producing a confirmed detection. Circles and squares indicate 50\% and 90\% detection thresholds.}
    \label{fig:detprob_venezuela_and_kashmir}
\end{figure}

Figure~\ref{fig:detprob_venezuela_and_kashmir} presents the
detection probability curves $P_{\mathrm{det}}(N, d)$ (Eq.~\ref{eq:pdet}) for
Region~H1 and Region~H5 respectively, under both the nanmean (left) and nanmax
(right) aggregation strategies.

\paragraph{Region~H1}
Under nanmean aggregation, $P_{\mathrm{det}}(7, d)$ reaches 100\% on
29~October~2025 for Region~H1, the earliest date among all constellation sizes.
$P_{\mathrm{det}}(3, d)$ reaches 100\% on 1~November~2025, and
$P_{\mathrm{det}}(1, d)$ on 8~December~2025. The ordering $N{=}7 \succ N{=}3
\succ N{=}1$ is monotone and consistent with the coverage argument above: each
additional satellite adds independent orbital passes that increase the
probability of two consecutive windows both exceeding the detection threshold.
All three curves reach 100\%, confirming that the Region~H1 RFI event is
detectable by any sub-constellation given sufficient observation time; the
multi-satellite advantage manifests exclusively in the speed of detection.
Table~\ref{tab:detection_summary} summarises the key detection timing milestones
for both regions.

\begin{table}[!t]
\centering
\caption{Detection probability milestones under nanmean aggregation with
$L{=}2$ streak confirmation. Dates correspond to the first day on which
$P_{\mathrm{det}}$ crosses each threshold. Dashes indicate the threshold was
not reached within the analysis period.}
\label{tab:detection_summary}
\setlength{\tabcolsep}{8pt}
\begin{tabular}{llcccc}
\toprule
\textbf{Region} & \textbf{Config} & \textbf{Final rate} &
    \textbf{$P \geq 50\%$} & \textbf{$P \geq 90\%$} &
    \textbf{$P = 100\%$} \\
\midrule
H1 & $N{=}1$ & 100\% & 29 Oct 2025 & 2 Nov 2025  & 2 Nov 2025  \\
   & $N{=}3$ & 100\% & 29 Oct 2025 & 29 Oct 2025 & 29 Oct 2025 \\
   & $N{=}7$ & 100\% & 15 Oct 2025 & 15 Oct 2025 & 15 Oct 2025 \\
\midrule
H5 & $N{=}1$ & 86\%  & 19 Oct 2025 & 24 Oct 2025 & ---         \\
   & $N{=}3$ & 100\% & 15 Oct 2025 & 19 Oct 2025 & 19 Oct 2025 \\
   & $N{=}7$ & 100\% & 15 Oct 2025 & 15 Oct 2025 & 15 Oct 2025 \\
\bottomrule
\end{tabular}
\end{table}

\paragraph{Region~H5}
The Region~H5 results reveal an important subtlety that distinguishes speed from
reliability in single-satellite monitoring. Under nanmean aggregation, the
$N{=}1$ curve rises earlier than $N{=}3$ and $N{=}7$. The median first
detection across the seven single-satellite combinations is 19~October, whereas
$N{=}3$ reaches 50\% on the same date. Yet, $P_{\mathrm{det}}(1, d)$ plateaus
at 86\% and never reaches 100\%. This apparent paradox is resolved by the
structure of CYGNSS orbital coverage over Region~H5.

Region~H5 is a geographically compact region ($\sim$$4^\circ \times 6^\circ$)
at mid-latitude, yielding the lowest slot coverage of any hotspot in the
analysis: 13\% for $N{=}1$ and 24\% for $N{=}7$.
With only 13\% of 90-min windows populated for a single satellite, most passes
over Region~H5 are isolated: the preceding and following windows are
\texttt{NaN}. The streak confirmation criterion (Eq.~\ref{eq:streak}) requires
two consecutive non-NaN windows both above threshold, a condition that is nearly
impossible to satisfy for the most unfavourably phased single-satellite
combinations, regardless of the actual RFI intensity. One of the seven CYGNSS
satellites never achieves two consecutive passes over Region~H5 within 90~min
throughout the entire 92-day analysis period; its $m^*(\mathcal{C})$ is
therefore permanently \texttt{NaN}, capping $P_{\mathrm{det}}(1, d)$ at
$6/7 \approx 86\%$.

The six remaining single-satellite combinations do achieve confirmation, and
several of them happen to do so early, on days when the satellite's orbital
phasing places two successive passes over Region~H5 within adjacent 90-min
windows during an active jamming period. This orbital coincidence produces the
fast initial rise of the $N{=}1$ curve. However, it is precisely a coincidence:
a single satellite that happens to be in the right orbital phase at the right
time, rather than a systematic detection capability. The 86\% reliability
ceiling exposes this: 14\% of single-satellite deployments would never confirm
the event at all.

The contrast with $N{=}3$ and $N{=}7$ is instructive. At $N{=}3$, the
merged series benefits from three independently phased satellites; the
probability that all three fail to produce two consecutive populated windows
simultaneously is much lower, and $P_{\mathrm{det}}(3, d)$ reaches 100\% by
19~October. At $N{=}7$, coverage is sufficient that consecutive slots are
almost always populated by different satellites, the streak requirement is
routinely satisfied, and the full ensemble detects by 15~October.

Under nanmax aggregation, the ordering reverses cleanly: $N{=}7$ detects
earliest at all confidence levels, followed by $N{=}3$ then $N{=}1$, and all
curves reach 100\%. This is the expected behaviour when the aggregation
preserves the strongest available signal rather than averaging it: more
satellites provide more chances for at least one satellite to catch the RFI
at peak intensity during any given window, and the single-window confirmation
threshold is more easily satisfied in proportion to the number of independent
sensors.

The Region~H1 and Region~H5 results together illustrate the two complementary
dimensions of multi-satellite advantage in persistence monitoring: for a
strong, geographically extended source such as Region~H1, additional satellites
accelerate detection without affecting ultimate reliability; for a marginal,
geographically compact source such as Region~H5, additional satellites are
necessary not only to reduce detection latency but to guarantee that detection
occurs at all.

\section{Discussion}
\label{sec:discussion}

Below, we provide a discussion of the four analytical dimensions examined
in this study: temporal responsiveness, spatial coverage, geometric
coherence, and persistence monitoring.

\subsection{Temporal Analysis}

The temporal coverage results demonstrate a clear benefit of constellation
scaling across all three metrics. Detection latency improves monotonically
from $N=1$ to $N=7$, with the full constellation achieving a $4.7\times$
reduction in median time-to-first-detection (48~minutes versus 3~hours
46~minutes). The diminishing returns observed between $N=3$, $N=5$, and
$N=7$ in the 90th-percentile latency suggest that a subset constellation
already captures the majority of geographically accessible hotspots, with
the remaining hard cases governed by coverage geometry rather than
satellite count. For transient emissions, the constellation advantage is
largest for the shortest bursts: the $5.8\times$ improvement in 5-minute
POI ($N=7$: 11.5\% versus $N=1$: 2.0\%) reflects the near-random nature
of short-burst interception, where additional satellites provide
statistically independent coverage opportunities. Beyond four hours, the
curves converge as the emission window becomes long enough for most
orbital configurations to guarantee at least one overpass.

The revisit results warrant comparison with the published CYGNSS mission
specifications, which report a median revisit time of approximately
2.8~hours and a mean of 7~hours for the full 8-satellite
constellation~(\cite{cygnss_brochure}). Our $N=7$ results yield a median
of 1~hour 49~minutes and a mean of 5~hours 19~minutes, broadly consistent
with and somewhat more favourable than the published figures. This is
physically expected: the published statistics characterise revisit for
arbitrary points across the coverage band, whereas our hotspots are
concentrated at low-to-mid latitudes over land, which receive denser
ground track coverage at the 35$^\circ$ orbital inclination. The agreement
nonetheless validates the revisit methodology. The large disparity between
median and mean across all configurations, most pronounced for $N=1$
(23~hours 9~minutes versus 27~hours 37~minutes), reflects the heavy-tailed
inter-pass gap distribution driven by periods of unfavourable orbital
geometry, an intrinsic property of low-Earth orbit constellations that
underscores the importance of reporting both statistics when characterising
surveillance cadence.

\subsection{Spatial Coverage Analysis}

The spatial coverage analysis reveals that constellation expansion yields
diminishing returns beyond N=5. While the transition from single-satellite to
three satellites delivers substantial gains in coverage fraction (37.9\% to
55.0\% at 6 hours) and median revisit time (5.8 hours to approximately 2
hours), additional satellites contribute marginal area coverage improvements.
This suggests constellation design should balance performance against mission
costs.

The persistent latitudinal gradient reflects an intrinsic limitation of the
35° inclined orbit. Equatorial regions receive two to three times more frequent
coverage than mid-latitude zones approaching orbital turnpoints, creating
geographic bias in temporal sampling. The 214 grid cells (0.2\%) that remain
unvisited over 48 hours concentrate in narrow longitudinal bands where orbital
phasing creates temporary gaps; extending analysis to 72 hours yields 99.97\%
coverage.

The 1-hour median revisit time achieved by the seven-satellite constellation
represents a critical threshold for operational response. Aviation safety
protocols generally require notification within 2 to 4 hours of interference
detection to enable countermeasures. The full constellation provides 90th
percentile revisit of 1.5 hours, supporting both initial detection and
persistence verification. In contrast, the single-satellite 10.5-hour 90th
percentile places most re-observations outside actionable response windows.

Combined with the sub-2-hour detection latency demonstrated in
Section~\ref{sec:temporal}, these results confirm that effective operational
GNSS interference monitoring requires multi-satellite architectures.
Single-satellite systems, while valuable for mapping and historical analysis,
cannot provide the temporal sampling density required for real-time threat
response.

\subsection{Geometric Coverage and Spatial Coherence}

The spatial coherence results establish a dimension of multi-satellite advantage
that is qualitatively distinct from the temporal and spatial coverage benefits
examined in preceding sections. Whereas detection latency and revisit time
are properties of when a satellite happens to overfly a candidate location,
the NRV quantifies whether the accumulated observations are sufficient to
characterise the spatial structure of the source, a prerequisite for
distinguishing a genuine point emitter from a residual sampling artefact
or a diffuse background elevation.

The elevated NRV values of 0.754 for Region~H3 and 0.793 for Region~H4 under
$N=1$ should not be interpreted as evidence that no RFI source is present, but
rather that a single satellite lacks the angular diversity of observations
required to resolve the source footprint. Observations concentrated in diagonal
orbital stripes with inter-swath voids of several hundred kilometres cannot
confirm spatial coherence even when a genuine source exists. Each additional
satellite fills these voids from a distinct orbital phase, progressively
isotropising the spatial sample distribution and reducing the residuals that
anisotropic sampling would otherwise produce.

The difference in absolute NRV levels between the two regions reflects a
genuine source property: the broader Region~H4 interference complex, spanning
approximately 2463~km in effective FWHM across a multi-emitter corridor, is
intrinsically harder to characterise with a single Gaussian template than the
more concentrated Region~H3 cluster at 1423~km. The asymptotic NRV floor of
approximately 0.35 for Region~H4 at $N=7$ accordingly reflects irreducible
model mismatch rather than insufficient sampling density.

The steepest NRV improvements in both regions occur between $N=1$ and $N=3$,
consistent with the pattern observed for temporal and spatial metrics in
preceding sections, and reinforcing the view that three satellites constitute
a qualitatively important architectural threshold below which the system
operates in a fundamentally information-limited regime. Beyond this threshold,
a multi-satellite constellation can certify the spatial integrity of a
candidate detection through cross-track corroboration, providing an
independent confirmation channel that complements temporal persistence
analysis and reducing the probability that a spatially incoherent false
positive propagates into an actionable detection report.

\subsection{Persistence Monitoring and Activity Pattern Analysis}

The persistence monitoring results establish two qualitatively distinct
dimensions of multi-satellite advantage that complement the temporal and
spatial coverage benefits examined in preceding sections. The first is
detection latency: for Region~H1, the $N{=}7$ constellation confirms the
interference onset on 29~October~2025, a date independently corroborated by
open-source reports published two days later documenting anomalous GPS
interference levels around Trinidad and Tobago and western
Venezuela~(\cite{trinidadexpress2025,gatechecked2025}), including a Notice to
Airmen issued by the Civil Aviation Authority of Trinidad and Tobago and
erroneous flight paths recorded on commercial tracking platforms. The
$N{=}1$ detection date of 8~December falls 39 days after this verified onset.
This gap is not a consequence of insufficient signal strength or detector
miscalibration; the jammer was documentably active on 29~October. It is a
structural consequence of the streak confirmation requirement interacting with
single-satellite orbital sparsity: at 2--3 passes per day over a compact
bounding box, consecutive 90-min slots are nearly always separated by multi-hour
gaps that reset the streak counter, regardless of the underlying RFI state.
The $N{=}7$ constellation avoids this failure mode because independent orbital
phases ensure that consecutive slots are populated by different satellites,
decoupling streak satisfaction from any single satellite's revisit geometry.

The second dimension is detection reliability rather than speed, and it emerges
most clearly in the Region~H5 results. The $N{=}1$ detection probability
plateaus at 86\%, meaning that one of the seven possible single-satellite
deployments would fail to confirm the event over the entire analysis period.
This ceiling is not recoverable by extending the observation window or
relaxing the detection threshold without a proportional increase in false
alarms: the unfavourably phased satellite simply never achieves two consecutive
populated windows over the $4^\circ\times6^\circ$ bounding box. At $N{=}3$
this ceiling disappears entirely, providing a quantitative argument for a
minimum constellation size of three whenever compact mid-latitude targets must
be monitored with guaranteed detection.

The comparison between nanmean and nanmax aggregation reveals a tradeoff
between sensitivity and specificity that has direct implications for
operational system design. Nanmax, by preserving the strongest single-satellite
observation at each window, produces earlier detections and eliminates the
reliability ceiling, but does so at the cost of an elevated susceptibility to
geometrically induced transients that the present analysis has not
systematically characterised. Nanmean with streak confirmation, by contrast,
requires the ensemble average to remain elevated across two successive windows,
a condition that suppresses isolated spike detections but whose
29~October detection date aligns with the independently verified interference
onset. The two strategies are therefore better understood as complementary
layers in a detection pipeline than as competing alternatives: nanmax provides
a sensitive first-pass screening that exploits the full spatial diversity of
the constellation, while nanmean streak confirmation provides the specificity
required before an operational alert is issued.

\section{Conclusion}
\label{sec:conclusion}

This paper evaluated GNSS-R RFI detection performance as a function of
constellation size across four dimensions, using real CYGNSS
observations over a three months period.
The full seven-satellite constellation reduces median detection latency by
$4.7\times$, raises 5-minute emission intercept probability from 2.0\% to
11.5\%, and compresses median revisit time from 6 hours to a less than 2 hours.
Spatial coherence analysis, quantified through Normalised Residual Variance,
demonstrates that a single satellite leaves up to 72\% of a source's spatial
signature unresolved, a dimension of detection confidence that temporal
metrics alone cannot capture. In persistence monitoring, the full
constellation confirmed the Region~H1 interference onset 39 days ahead of
any single-satellite deployment, on a date independently corroborated by
published aviation notices, and eliminated a hard 86\% detection reliability
ceiling over Region~H5 that is unrecoverable through threshold tuning at the
single-satellite level.

The steepest improvements across all metrics occur between $N=1$ and $N=3$,
establishing a three-satellite minimum as the qualitative threshold below
which orbital geometry rather than signal physics governs detection failure.
For space-based GNSS-R RFI detection and monitoring architectures, this
suggests that constellation sizing should account for detection reliability
over compact mid-latitude targets and not be driven by aggregate coverage
fraction alone.

The present analysis is confined to $\pm40°$ latitude and a single three months
period. Extending the framework to higher inclination architectures,
characterising false alarm rates systematically, and integrating
the sensitivity enhancement analysis into a unified detection pipeline
represent the natural next steps toward an operational space-based RFI
monitoring system.

\section*{Acknowledgements}

This work was supported in part by the Tier 1 Canada Research Chair Program. The authors acknowledge NASA's Physical Oceanography Distributed Active Archive Center for providing access to CYGNSS data products. The OPeNDAP service and its efficient data access substantially facilitated this research. We thank the CYGNSS mission team at the University of Michigan for their continued operation of this valuable scientific resource. We also thank Naron Phou and Yann Picard from Northstar Earth \& Space for their valuable contributions and review of this work. Finally, we acknowledge \cite{chew2023rfi} for inspiring the idea behind this paper.

\bibliographystyle{apalike}
\bibliography{citations}

\end{document}